\documentclass[iop]{emulateapj}
\usepackage{graphicx}
\usepackage{IEEEtrantools}
\usepackage{hyperref}

\newcommand{\bdv}[1]{\mbox{\boldmath$#1$}}
\newcommand{\dl}{\ensuremath{D_{\ell}}}
\newcommand{\dpl}{\ensuremath{D_{p}}}
\newcommand{\ds}{\ensuremath{D_{s}}}
\newcommand{\fb}{\ensuremath{F_{b}}}
\newcommand{\fl}{\ensuremath{F_{\ell}}}
\newcommand{\fsource}{\ensuremath{F_{s}}}
\newcommand{\ml}{\ensuremath{M_{\ell}}}
\newcommand{\mstar}{\ensuremath{M_{*}}}
\newcommand{\mpl}{\ensuremath{M_{p}}}
\newcommand{\murel}{\ensuremath{\mu_{\rm rel}}}
\newcommand{\piE}{\ensuremath{\pi_{\rm E}}}
\newcommand{\pirel}{\ensuremath{\pi_{\rm rel}}}
\newcommand{\tE}{\ensuremath{t_{\rm E}}}
\newcommand{\thetaE}{\ensuremath{\theta_{\rm E}}}
\newcommand{\thetastar}{\ensuremath{\theta_{*}}}
\newcommand{\tzero}{\ensuremath{t_{\rm 0}}}
\newcommand{\uzero}{\ensuremath{u_{\rm 0}}}

\newcommand{\cn}{{C9}}
\newcommand{\cz}{{Campaign 0}}
\newcommand{\kepler}{{\it Kepler}}
\newcommand{\kt}{{\it K2}}
\newcommand{\ktcn}{{\it K2}C9}
\newcommand{\spitzer}{{\it Spitzer}}
\newcommand{\wfirst}{{\it WFIRST}}

\begin{document}

\begin{turnpage}
\end{turnpage}

\title{
Campaign 9 of the {\kt} Mission:\\
Observational Parameters, Scientific Drivers, and Community Involvement\\
for a Simultaneous Space- and Ground-based Microlensing Survey
}

\author{
%%%%%%%%%%%%%%%%%%%%%%%%%
%%% Microlensing Science Team
%%%%%%%%%%%%%%%%%%%%%%%%%
Calen B. Henderson\altaffilmark{1,A},   %%% Lead author; MST member; PI for IRTF; leading ANDICAM; Co-I for Keck; assisting with UKIRT
Rados{\l}aw Poleski\altaffilmark{2,3},   %%% Pixel selection; Co-I for IRTF
Matthew Penny\altaffilmark{2,B},   %%% MST member; PI SkyMapper; leading DECam; Co-I for Keck, IRTF, CFHT, VST
Rachel A. Street\altaffilmark{4},   %%% MST member; RoboNet; Leading LCOGT; PI for LT (?); Co-I for VST
David P. Bennett\altaffilmark{5},\\   %%% MST member; MOA; Co-I for Subaru
David W. Hogg\altaffilmark{6,7},   %%% MST member
B. Scott Gaudi\altaffilmark{2}\\   %%% Ex Officio MST member; Co-I on MP MST proposal; Co-I for Keck, IRTF
({\kt} Campaign 9 Microlensing Science Team),\\
and\\
%%%%%%%%%%%%%%%%%%%%%%%%%
%%% Other key personnel (Radek+NASA)
%%%%%%%%%%%%%%%%%%%%%%%%%
W. Zhu\altaffilmark{2},   %%% Co-I for IRTF, CFHT
T. Barclay\altaffilmark{8},   %%% K2 Team
G. Barentsen\altaffilmark{8},   %%% K2 Team
S. B. Howell\altaffilmark{8},   %%% K2 Team
F. Mullally\altaffilmark{8},\\   %%% K2 Team
and\\
%%%%%%%%%%%%%%%%%%%%%%%%%
%%% OGLE
%%%%%%%%%%%%%%%%%%%%%%%%%
A. Udalski\altaffilmark{3},   %%% Co-I on APG MST proposal; PI for OGLE
M. K. Szyma{\'n}ski\altaffilmark{3},   %%% OGLE
J. Skowron\altaffilmark{3},   %%% OGLE
P. Mr{\'o}z\altaffilmark{3},   %%% OGLE
S. Koz{\l}owski\altaffilmark{3},   %%% OGLE
{\L}. Wyrzykowski\altaffilmark{3},   %%% OGLE
P. Pietrukowicz\altaffilmark{3},\\   %%% OGLE
I. Soszy{\'n}ski\altaffilmark{3},   %%% OGLE
K. Ulaczyk\altaffilmark{3},   %%% OGLE
M. Pawlak\altaffilmark{3}\\   %%% OGLE
(The OGLE Project),\\
and\\
%%%%%%%%%%%%%%%%%%%%%%%%%
%%% MOA
%%%%%%%%%%%%%%%%%%%%%%%%%
T. Sumi\altaffilmark{9},   %%% Co-I on DPB MST proposal; MOA; Co-I for Keck, Subaru
F. Abe\altaffilmark{10},   %%% MOA
Y. Asakura\altaffilmark{9},   %%% MOA? email from Taka [27/Dec/2015]
R. K. Barry\altaffilmark{5},   %%% MOA? email from Taka [27/Dec/2015]
A. Bhattacharya\altaffilmark{11},   %%% MOA
I. A. Bond\altaffilmark{12},   %%% Co-I on DPB's MST proposal; MOA
M. Donachie\altaffilmark{13},   %%% MOA
M. Freeman\altaffilmark{13},\\   %%% MOA
A. Fukui\altaffilmark{14},   %%% MOA; Co-I for Subaru
Y. Hirao\altaffilmark{9},   %%% MOA
Y. Itow\altaffilmark{10},   %%% MOA
N. Koshimoto\altaffilmark{9},   %%% MOA
M. C. A. Li\altaffilmark{13},   %%% MOA
C. H. Ling\altaffilmark{12},   %%% MOA
K. Masuda\altaffilmark{10},   %%% MOA
Y. Matsubara\altaffilmark{10},\\   %%% MOA
Y. Muraki\altaffilmark{10},   %%% MOA
M. Nagakane\altaffilmark{9},   %%% MOA
K. Ohnishi\altaffilmark{15},   %%% MOA
H. Oyokawa\altaffilmark{9},   %%% MOA? email from Taka [27/Dec/2015]
N. Rattenbury\altaffilmark{13},   %%% MOA
To. Saito\altaffilmark{16},   %%% MOA
A. Sharan\altaffilmark{13},\\   %%% MOA
D. J. Sullivan\altaffilmark{17},   %%% MOA
P. J. Tristram\altaffilmark{18},   %%% MOA
A. Yonehara\altaffilmark{19}\\   %%% MOA
(The MOA Collaboration),\\
and\\
%%%%%%%%%%%%%%%%%%%%%%%%%
%%% RoboNet
%%%%%%%%%%%%%%%%%%%%%%%%%
E. Bachelet\altaffilmark{4},   %%% PI for WIYN; RoboNet; Co-I for VST
D. M. Bramich\altaffilmark{20},   %%% RoboNet
A. Cassan\altaffilmark{21},   %%% RoboNet; Co-I for VST
M. Dominik\altaffilmark{22},   %%% Co-I on RAS MST proposal; RoboNet; Co-I for VST
R. Figuera Jaimes\altaffilmark{22},   %%% Co-I on RAS MST proposal; RoboNet
K. Horne\altaffilmark{22},   %%% Co-I on RAS MST proposal; RoboNet
M. Hundertmark\altaffilmark{23},\\   %%% Co-I on RAS MST proposal; PI for VST; RoboNet
S. Mao\altaffilmark{24,25,26},   %%% PI for CFHT; RoboNet
C. Ranc\altaffilmark{21},   %%% RoboNet
R. Schmidt\altaffilmark{27},   %%% RoboNet
C. Snodgrass\altaffilmark{28},   %%% RoboNet; Co-I for VST
I. A. Steele\altaffilmark{29},   %%% RoboNet
Y. Tsapras\altaffilmark{27},   %%% Co-I on RAS MST proposal; RoboNet; Co-I for VST
J. Wambsganss\altaffilmark{27}\\   %%% RoboNet
(The RoboNet Project),\\
and\\
%%%%%%%%%%%%%%%%%%%%%%%%%
%%% MiNDSTEp
%%%%%%%%%%%%%%%%%%%%%%%%%
V. Bozza\altaffilmark{30,31},   %%% MiNDSTEp; Co-I on RAS's MST proposal; real-time modeler
M. J. Burgdorf\altaffilmark{32},   %%% MiNDSTEp
U. G. J{\o}rgensen\altaffilmark{23},   %%% MiNDSTEp; Co-I for VST
S. Calchi Novati\altaffilmark{30,33,34},   %%% MiNDSTEp
S. Ciceri\altaffilmark{35},   %%% MiNDSTEp
G. D'Ago\altaffilmark{34},\\   %%% MiNDSTEp
D. F. Evans\altaffilmark{36},   %%% MiNDSTEp
F. V. Hessman\altaffilmark{37},   %%% MiNDSTEp
T. C. Hinse\altaffilmark{38},   %%% MiNDSTEp
T.-O. Husser\altaffilmark{37},   %%% MiNDSTEp
L. Mancini\altaffilmark{35},   %%% MiNDSTEp
A. Popovas\altaffilmark{23},   %%% MiNDSTEp
M. Rabus\altaffilmark{39},\\   %%% MiNDSTEp
S. Rahvar\altaffilmark{40},   %%% MiNDSTEp
G. Scarpetta\altaffilmark{30,34},   %%% MiNDSTEp
J. Skottfelt\altaffilmark{41,23},   %%% MiNDSTEp
J. Southworth\altaffilmark{36},   %%% MiNDSTEp
E. Unda-Sanzana\altaffilmark{42}\\   %%% MiNDSTEp
(The MiNDSTEp Team),\\
and\\
%%%%%%%%%%%%%%%%%%%%%%%%%
%%% Engineering Team
%%%%%%%%%%%%%%%%%%%%%%%%%
S. T. Bryson\altaffilmark{8},   %%% NASA Ames
D. A. Caldwell\altaffilmark{8},   %%% NASA Ames
M. R. Haas\altaffilmark{8},   %%% NASA Ames
K. Larson\altaffilmark{43},   %%% Ball Aerospace
K. McCalmont\altaffilmark{43},   %%% Ball Aerospace
M. Packard\altaffilmark{44},   %%% UC Boulder
C. Peterson\altaffilmark{43},\\   %%% Ball Aerospace
D. Putnam\altaffilmark{43},   %%% Ball Aerospace
L. Reedy\altaffilmark{44},   %%% UC Boulder
S. Ross\altaffilmark{43},   %%% Ball Aerospace
J. E. Van Cleve\altaffilmark{8}\\   %%% NASA Ames
({\ktcn} Engineering Team),\\
and\\
%%%%%%%%%%%%%%%%%%%%%%%%%
%%% Co-Is and associated scientists
%%%%%%%%%%%%%%%%%%%%%%%%%
R. Akeson\altaffilmark{33},   %%% ExoFOP
V. Batista\altaffilmark{21},   %%% Email from JP [23/Dec]; Co-I for Subaru
J.-P. Beaulieu\altaffilmark{21},   %%% Co-I on DPB's MST proposal; Co-I for Subaru
C. A. Beichman\altaffilmark{45,1,33},   %%% Co-I on CBH's MST proposal; PI for Keck; Co-I for IRTF
G. Bryden\altaffilmark{1},   %%% Co-I on IRTF
D. Ciardi\altaffilmark{33},   %%% ExoFOP
A. Cole\altaffilmark{46},\\   %%% Email from JP [23/Dec]
C. Coutures\altaffilmark{21},   %%% Email from JP [25/Dec]
D. Foreman-Mackey\altaffilmark{47,B},   %%% Co-I on MST proposals for both DWH and MP
P. Fouqu\'{e}\altaffilmark{48},   %%% Co-I for CFHT
M. Friedmann\altaffilmark{49},   %%% Co-I for Wise
C. Gelino\altaffilmark{33},   %%% Co-I for Keck
S. Kaspi\altaffilmark{49},   %%% Co-I for Wise
E. Kerins\altaffilmark{50},\\   %%% Co-I for VST
H. Korhonen\altaffilmark{23},   %%% Co-I for VST
D. Lang\altaffilmark{51},   %%% Co-I on DWH MST proposal
C.-H. Lee\altaffilmark{52},   %%% PI for Subaru
C. H. Lineweaver\altaffilmark{53},   %%% Co-I for SkyMapper
D. Maoz\altaffilmark{49},   %%% Co-I for Wise
J.-B. Marquette\altaffilmark{21},   %%% Email from JP [23/Dec]
F. Mogavero\altaffilmark{21},\\   %%% Email from JP [23/Dec]
J. C. Morales\altaffilmark{54},   %%% Email from JP [25/Dec]
D. Nataf\altaffilmark{53},   %%% Co-I for SkyMapper
R. W. Pogge\altaffilmark{2},   %%% Co-I on CBH MST proposal
A. Santerne\altaffilmark{55},   %%% Email from JP [25/Dec]
Y. Shvartzvald\altaffilmark{1,A},   %%% Leading UKIRT; Co-I for IRTF
D. Suzuki\altaffilmark{5},   %%% Co-I on DPB's MST proposal; Co-I for Subaru
M. Tamura\altaffilmark{56,57,58},\\   %%% Co-I for Keck
P. Tisserand\altaffilmark{21},   %%% Co-I for SkyMapper
D. Wang\altaffilmark{6}\\   %%% Co-I on DWH MST proposal
\vspace{-1.00mm}
}
%%%%%%%%%%%%%%%%%%%%%%%%%
%%% Institutional affiliations
%%%%%%%%%%%%%%%%%%%%%%%%%
\altaffiltext{1}{Jet Propulsion Laboratory, California Institute of Technology, 4800 Oak Grove Drive, Pasadena, CA 91109, USA}
\altaffiltext{2}{Department of Astronomy, Ohio State University, 140 W. 18th Ave., Columbus, OH  43210, USA}
\altaffiltext{3}{Warsaw University Observatory, Al. Ujazdowskie 4, 00-478 Warszawa, Poland}
\altaffiltext{4}{Las Cumbres Observatory Global Telescope Network, 6740 Cortona Drive, suite 102, Goleta, CA 93117, USA}
\altaffiltext{5}{Laboratory for Exoplanets and Stellar Astrophysics, NASA Goddard Space Flight Center, Greenbelt, MD 20771, USA}
\altaffiltext{6}{Center for Cosmology and Particle Physics, Department of Physics, New York University, 4 Washington Pl., room 424, New York, NY 10003, USA}
\altaffiltext{7}{Center for Data Science, New York University, 726 Broadway, 7th Floor, New York, NY 10003, USA}
\altaffiltext{8}{NASA Ames Research Center, Moffett Field, CA 94035}
\altaffiltext{9}{Department of Earth and Space Science, Graduate School of Science, Osaka University, Toyonaka, Osaka 560-0043, Japan}
\altaffiltext{10}{Institute for Space-Earth Environmental Research, Nagoya University, Nagoya 464-8601, Japan}
\altaffiltext{11}{Department of Physics, University of Notre Dame, Notre Dame, IN 46556, USA}
\altaffiltext{12}{Institute of Information and Mathematical Sciences, Massey University, Private Bag 102-904, North Shore Mail Centre, Auckland, New Zealand}
\altaffiltext{13}{Department of Physics, University of Auckland, Private Bag 92019, Auckland, New Zealand}
\altaffiltext{14}{Okayama Astrophysical Observatory, National Astronomical Observatory of Japan, 3037-5 Honjo, Kamo-gata, Asakuchi, Okayama 719-0232, Japan}
\altaffiltext{15}{Nagano National College of Technology, Nagano 381-8550, Japan}
\altaffiltext{16}{Tokyo Metropolitan College of Aeronautics, Tokyo 116-8523, Japan}
\altaffiltext{17}{School of Chemical and Physical Sciences, Victoria University, Wellington, New Zealand}
\altaffiltext{18}{Mt. John University Observatory, P.O. Box 56, Lake Tekapo 8770, New Zealand}
\altaffiltext{19}{Department of Physics, Faculty of Science, Kyoto Sangyo University, 603-8555 Kyoto, Japan}
\altaffiltext{20}{Qatar Environment and Energy Research Institute (QEERI), HBKU, Qatar Foundation, Doha, Qatar}
\altaffiltext{21}{Sorbonne Universit{\'e}s, UPMC Univ Paris 6 et CNRS, UMR 7095, Institut d'Astrophysique de Paris, 98 bis bd Arago, 75014 Paris, France}
\altaffiltext{22}{SUPA, University of St Andrews, School of Physics \& Astronomy, North Haugh, St. Andrews KY16 9SS, United Kingdom}
\altaffiltext{23}{Niels Bohr Institute \& Centre for Star and Planet Formation, University of Copenhagen, {\O}ster Voldgade 5, 1350 Copenhagen, Denmark}
\altaffiltext{24}{Department of Physics and Center for Astrophysics, Tsinghua University, Haidian District, Beijing 100084, China}
\altaffiltext{25}{National Astronomical Observatories, 20A Datun Road, Chinese Academy of Sciences, Beijing 100012, China}
\altaffiltext{26}{Jodrell Bank Centre for Astrophysics, School of Physics and Astronomy, University of Manchester, Alan Turing Building, Oxford Road, Manchester M13 9PL, UK}
\altaffiltext{27}{Astronomisches Rechen-Institut, Zentrum f{\"u}r Astronomie der Universit{\"a}t Heidelberg (ZAH), 69120 Heidelberg, Germany}
\altaffiltext{28}{Planetary and Space Sciences, Department of Physical Sciences, The Open University, Milton Keynes, MK7 6AA, UK}
\altaffiltext{29}{Astrophysics Research Institute, Liverpool John Moores University, Liverpool CH41 1LD, UK}
\altaffiltext{30}{Dipartimento di Fisica ``E.R. Caianiello", Universit{\`a} di Salerno, Via Giovanni Paolo II 132, I-84084 Fisciano (SA), Italy}
\altaffiltext{31}{Istituto Nazionale di Fisica Nucleare, Sezione di Napoli, Napoli, Italy}
\altaffiltext{32}{Meteorologisches Institut, Universit{\"a}t Hamburg, Bundesstra\ss{}e 55, 20146 Hamburg, Germany}
\altaffiltext{33}{NASA Exoplanet Science Institute, California Institute of Technology, 770 S. Wilson Ave., Pasadena, CA 91125}
\altaffiltext{34}{Istituto Internazionale per gli Alti Studi Scientifici (IIASS), Via G. Pellegrino 19, 84019 Vietri sul Mare (SA), Italy}
\altaffiltext{35}{Max Planck Institute for Astronomy, K{\"o}nigstuhl 17, 69117 Heidelberg, Germany}
\altaffiltext{36}{Astrophysics Group, Keele University, Staffordshire, ST5 5BG, UK}
\altaffiltext{37}{Institut f{\"u}r Astrophysik, Georg-August-Universit{\"a}t, Friedrich-Hund-Platz 1, 37077 G{\"o}ttingen, Germany}
\altaffiltext{38}{Korea Astronomy \& Space Science Institute, 776 Daedukdae-ro, Yuseong-gu, 305-348 Daejeon, Republic of Korea}
\altaffiltext{39}{Instituto de Astrof{\'i}sica, Facultad de F{\'i}sica, Pontificia Universidad Cat{\'o}lica de Chile, Av. Vicu\~na Mackenna 4860, 7820436 Macul, Santiago, Chile}
\altaffiltext{40}{Department of Physics, Sharif University of Technology, PO Box 11155-9161 Tehran, Iran}
\altaffiltext{41}{Centre for Electronic Imaging, Department of Physical Sciences, The Open University, Milton Keynes, MK7 6AA, UK}
\altaffiltext{42}{Unidad de Astronom{\'i}a, Fac. de Ciencias B{\'a}sicas, Universidad de Antofagasta, Avda. U. de Antofagasta 02800, Antofagasta, Chile}
\altaffiltext{43}{Ball Aerospace \& Technologies, Boulder, CO, 80301}
\altaffiltext{44}{Laboratory for Atmospheric and Space Physics, University of Colorado at Boulder, Boulder, CO, 80303}
\altaffiltext{45}{Infrared Processing and Analysis Center, California Institute of Technology, Pasadena CA 91125}
\altaffiltext{46}{School of Physical Sciences, University of Tasmania, Private Bag 37 Hobart, Tasmania 7001 Australia}
\altaffiltext{47}{Astronomy Department, University of Washington, Seattle, WA 98195, USA}
\altaffiltext{48}{CFHT Corporation 65-1238 Mamalahoa Hwy Kamuela, Hawaii 96743, USA}
\altaffiltext{49}{School of Physics and Astronomy, Tel-Aviv University, Tel-Aviv 69978, Israel}
\altaffiltext{50}{School of Physics and Astronomy, University of Manchester, Oxford Road, Manchester M13 9PL}
\altaffiltext{51}{Department of Astronomy and Astrophysics, University of Toronto, 50 St. George Street, Toronto, Ontario, Canada M5S 3H4}
\altaffiltext{52}{Subaru Telescope, National Astronomical Observatory of Japan, 650 North Aohoku Place, Hilo, HI 96720, USA}
\altaffiltext{53}{Research School of Astronomy and Astrophysics, Australian National University, Canberra, ACT 2611, Australia}
\altaffiltext{54}{Institut de Ci{\`e}ncies de l'Espai (CSIC-IEEC), Campus UAB, Carrer de Can Magrans s/n, 08193 Cerdanyola del Vall{\`e}s, Spain}
\altaffiltext{55}{Instituto de Astrof{\'i}sica e Ci{\^e}ncias do Espa\c{c}o, Universidade do Porto, CAUP, Rua das Estrelas, 4150-762 Porto, Portugal}
\altaffiltext{56}{Astrobiology Center, 2-21-1 Osawa, Mitaka, Tokyo, 181-8588, Japan}
\altaffiltext{57}{National Astronomical Observatory of Japan, 2-21-1 Osawa, Mitaka, Tokyo, 181-8588, Japan}
\altaffiltext{58}{Department of Astronomy, The University of Tokyo, 7-3-1 Hongo, Bunkyo-ku, Tokyo, 113-0033, Japan}
%%%%%%%%%%%%%%%%%%%%%%%%%
%%% Fellowship affiliations
%%%%%%%%%%%%%%%%%%%%%%%%%
\altaffiltext{A}{NASA Postdoctoral Program Fellow}
\altaffiltext{B}{Sagan Fellow}
%%%%%%%%%%%%%%%%%%%%%%%%%
%%% Email address for corresponding author
%%%%%%%%%%%%%%%%%%%%%%%%%
\email{calen.b.henderson@jpl.nasa.gov}

%%%%%%%%%%%%%%%%%%%%%%%%%%%%%%%%%%%%%%%%%%%%%%%%%%
%%%
\begin{abstract}
%%%
%%%%%%%%%%%%%%%%%%%%%%%%%%%%%%%%%%%%%%%%%%%%%%%%%%

{\kt}'s Campaign 9 (\ktcn) will conduct a $\sim$3.7 deg$^{2}$ survey toward the Galactic bulge from 7/April through 1/July of 2016 that will leverage the spatial separation between {\kt} and the Earth to facilitate measurement of the microlens parallax {\piE} for $\gtrsim$127 microlensing events.
These will include several that are planetary in nature as well as many short-timescale microlensing events, which are potentially indicative of free-floating planets (FFPs).
These satellite parallax measurements will in turn allow for the direct measurement of the masses of and distances to the lensing systems.
In this white paper we provide an overview of the {\ktcn} space- and ground-based microlensing survey.
Specifically, we detail the demographic questions that can be addressed by this program, including the frequency of FFPs and the Galactic distribution of exoplanets, the observational parameters of {\ktcn}, and the array of resources dedicated to concurrent observations.
Finally, we outline the avenues through which the larger community can become involved, and generally encourage participation in {\ktcn}, which constitutes an important pathfinding mission and community exercise in anticipation of {\wfirst}.
\end{abstract}


%%%%%%%%%%%%%%%%%%%%%%%%%%%%%%%%%%%%%%%%%%%%%%%%%%
%%%
\section{{Introduction} \label{sec:intro}}
%%%
%%%%%%%%%%%%%%%%%%%%%%%%%%%%%%%%%%%%%%%%%%%%%%%%%%

Results from the {\kepler} Mission have revolutionized our understanding of the frequency and distribution of exoplanets that orbit close-in to their host stars.
To-date it has identified 4175 planet candidates \citep{mullally2015} and has confirmed 1039 as bona fide exoplanets\footnote{From \url{http://kepler.nasa.gov/}}.
These discoveries have led to a wealth of insights into exoplanet demographics, including the apparent ubiquity of small planets (e.g., \citealt{fressin2013}) and the occurrence rate and orbital architectures of systems with multiple transiting planet candidates (e.g., \citealt{fabrycky2014}), along the quest to measure $\eta_{\oplus}$.

The mechanical failure of the second of {\kepler}'s four reaction wheels in 2013 signaled an end to the primary mission but heralded the genesis of its extended {\kt} Mission, which is in the midst of a series of $\sim$80-day campaigns performing high-precision photometry for targets along the Ecliptic \citep{howell2014}.
Orienting the spacecraft to point along its velocity vector (+VV) allows {\kt}'s Campaign 9 (\ktcn) to observe toward the Galactic bulge while it is simultaneously visible from Earth, enabling the first microlensing survey from both the ground and from space.

In this white paper we detail the joint space- and ground-based microlensing survey enabled by {\ktcn}.
We begin with a brief overview of the geometric principles and observational implementation of the microlensing technique in \S \ref{sec:ulens_bkgrd}.
Then, in \S \ref{sec:science} we discuss the scientific questions to which {\ktcn} will provide access.
This is followed by a description of the observational parameters of {\ktcn} in \S \ref{sec:k2_parms}.
In \S \ref{sec:ground} we summarize the ground-based resources that will be employed concurrently with {\ktcn}, as well as their scientific goals.
We detail the goals and implementation of a 50-hour {\spitzer} program that will take simultaneous observations during the last 13 days of {\ktcn} in \S \ref{sec:spitzer}.
Finally, in \S \ref{sec:community} we focus on the channels through which the greater community can participate in this community-driven microlensing experiment.

%%%%%%%%%%%%%%%%%%%%%%%%%%%%%%%%%%%%%%%%%%%%%%%%%%
%%%
\section{Gravitational Microlensing Overview} \label{sec:ulens_bkgrd}
%%%
%%%%%%%%%%%%%%%%%%%%%%%%%%%%%%%%%%%%%%%%%%%%%%%%%%

In this section we provide a brief overview of the theoretical background and observational implementation of gravitational microlensing.
See \citet{gaudi2012} for a deeper exploration of the fundamental mechanics of lensing, particularly in the context of searches for exoplanets.
Readers who possess a foundational understanding of the technical details of microlensing should proceed to \S \ref{sec:science}.

%------------------------------------------------------------------------------------------------------------------------
\subsection{Lensing Geometry} \label{sec:ulens_geometry}
%------------------------------------------------------------------------------------------------------------------------

\begin{figure}
   \centerline{
      \includegraphics[width=9cm]{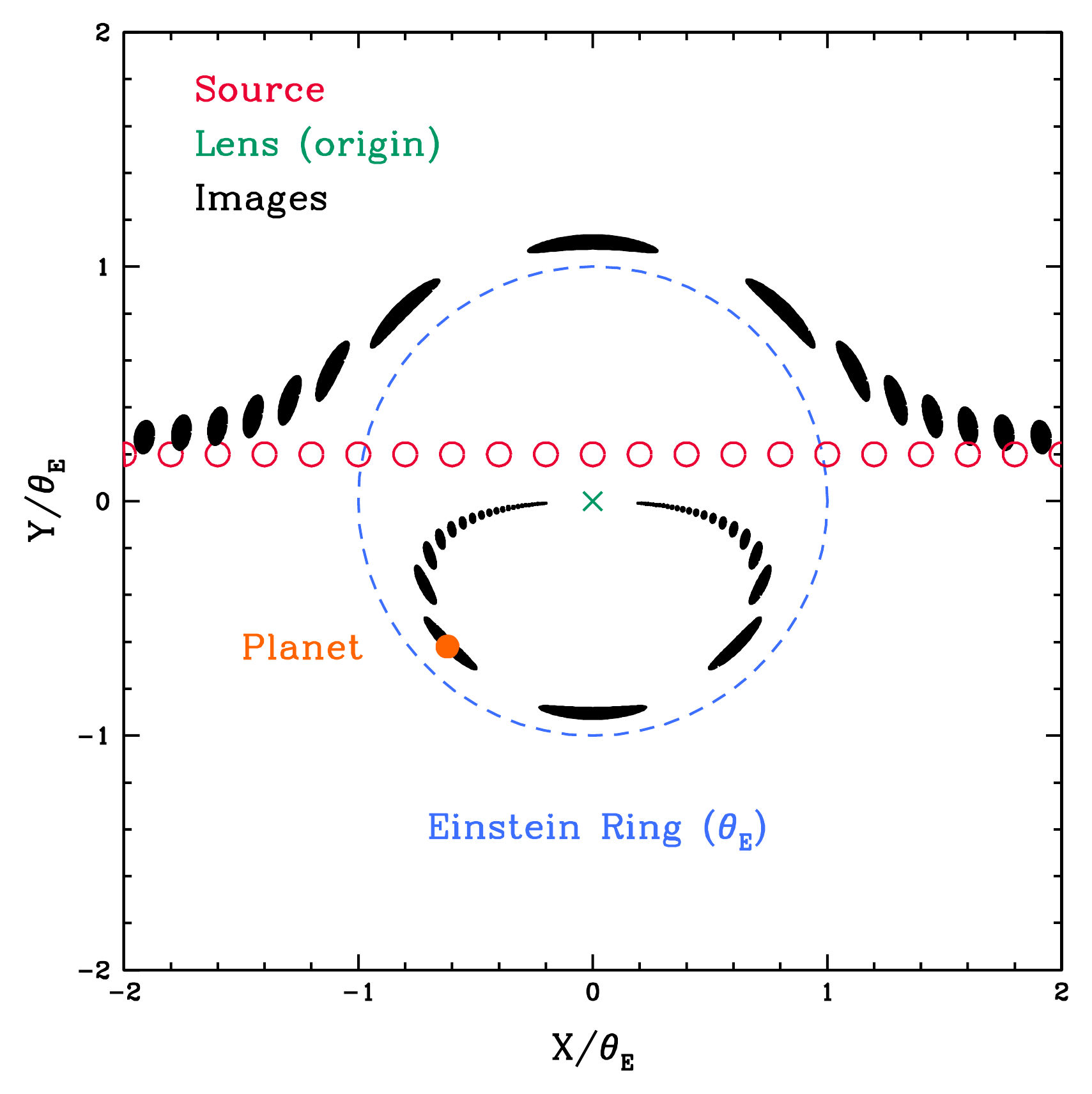}
   }
   \caption{
      \footnotesize{
         The face-on geometry of the temporal evolution of a microlensing event due to a single lensing mass.
         The green cross at the origin denotes the fixed lens position, the red open circles identify the trajectory of the source, and the filled black ellipses show the paths of the two images created during the event.
         In the case of exact observer-lens-source colinearity, the two images merge to create a circle with radius equal to {\thetaE}.
         The introduction of a second body, such as a planet marked by the filled orange circle, approximately coincident with one of the images will introduce additional magnification structure to the light curve.
         \vspace{1.0mm}
      }
   }
   \label{fig:ulens_geometry}
\end{figure}

A microlensing event occurs when the light from a background ``source'' star is magnified by the gravitational potential of an intervening foreground ``lens'' system in a way that is detectable by a given observer.
When describing the temporal evolution of an event, as is shown in Figure \ref{fig:ulens_geometry}, the coordinate system keeps the lensing body fixed at the origin such that all of the lens-source relative proper motion is encapsulated in the trajectory of the source.
The light from the source is split into two images that, in the case of perfect observer-lens-source colinearity, trace out the Einstein radius {\thetaE}, the angular scale for microlensing phenomena.

For a lensing system with total mass {\ml} the Einstein ring is defined as:
\begin{equation} \label{eq:thetaE}
   {\thetaE} \equiv \sqrt{\kappa{\ml}{\pirel}},~\pi_{\rm rel} = \pi_{\rm E}\theta_{\rm E} = {\rm AU}({\dl}^{-1} - {\ds}^{-1}),
\end{equation}
where $\kappa \equiv 4G/(c^{2}{\rm AU}) = 8.144~{\rm mas}/M_{\odot}$, $\pi_{\rm rel}$ is the relative lens-source parallax, and {\dl} and {\ds} are the distances to the lens and source, respectively.
Normalizing the relative lens-source parallax to {\thetaE} yields the microlens parallax {\piE}.
For typical microlensing surveys toward the Galactic bulge, {\thetaE} is of-order a milliarcsecond or smaller, meaning that the images of the source are not spatially resolved.

An event due to a single lensing mass leads to a light curve defined by three microlensing observables \citep{paczynski1986}.
The first is {\tzero}, the time of closest approach of the source to the lens.
Second is the impact parameter {\uzero}, which measures the angular distance of the closest approach of the source to the lens and is normalized to {\thetaE}.
Lastly, the Einstein crossing time {\tE} is defined via:
\begin{equation} \label{eq:tE}
   t_{\rm E} \equiv \frac{\thetaE}{\murel},
\end{equation}
where {\murel} is the relative lens-source proper motion.

Since the two images created during the event are not resolved, a given observer measures the total flux of the event.
For a single-lens microlensing event the observed flux $F$ at a time $t$ is given by:
\begin{equation} \label{eq:f_obs}
   F(t) = {\fsource}A(u) + {\fb},
\end{equation}
where {\fsource} is the flux of the source, {\fb} is the blend flux of all other stars that are not resolved, and $A(u)$ is the magnification of the point-like background source star:
\begin{equation} \label{eq:magnification}
   A(u) = \frac{u^{2} + 2}{u\sqrt{u^{2} + 4}}.
\end{equation}
Here $u$ is the angular separation of the lens and source at a given time $t$, normalized to {\thetaE}.

In some cases, higher-order effects are imprinted on the light curve.
Among these is $\rho$, the angular radius of the source star, {\thetastar}, normalized to {\thetaE}:
\begin{equation} \label{eq:rho}
   \rho \equiv \frac{\thetastar}{\thetaE}.
\end{equation}
Finite-source effects, caused by a value of $\rho$ that is comparable to {\uzero}, and also the higher-order microlens parallax {\piE}, alter the magnification structure of the light curve, causing it to deviate from a simple Paczynski curve.

If the lensing system contains an additional mass whose position is roughly coincident with that of one of the images at any point during the event, the additional gravitational potential introduced by the second body will distort the magnification structure of the event \citep{mao1991,gould1992b}.
In the case of a static two-body lensing system, such as a planet orbiting a host star, these perturbations allow for the measurement of three additional parameters.
The mass ratio $q$ of a lens comprised of a planet of mass {\mpl} and a star of mass {\mstar} is given by:
\begin{equation} \label{eq:q}
   q = \frac{\mpl}{\mstar}.
\end{equation}
The instantaneous projected angular separation of the two bodies, normalized to {\thetaE}, is denoted by $s$.
Finally, $\alpha$ gives the angle of the source trajectory relative to the star-planet binary axis.
See \citet{gould2000a} and \citet{skowron2011} for a more complete discussion of microlensing notation conventions.

\begin{figure*}
   \centerline{
      \includegraphics[width=18cm]{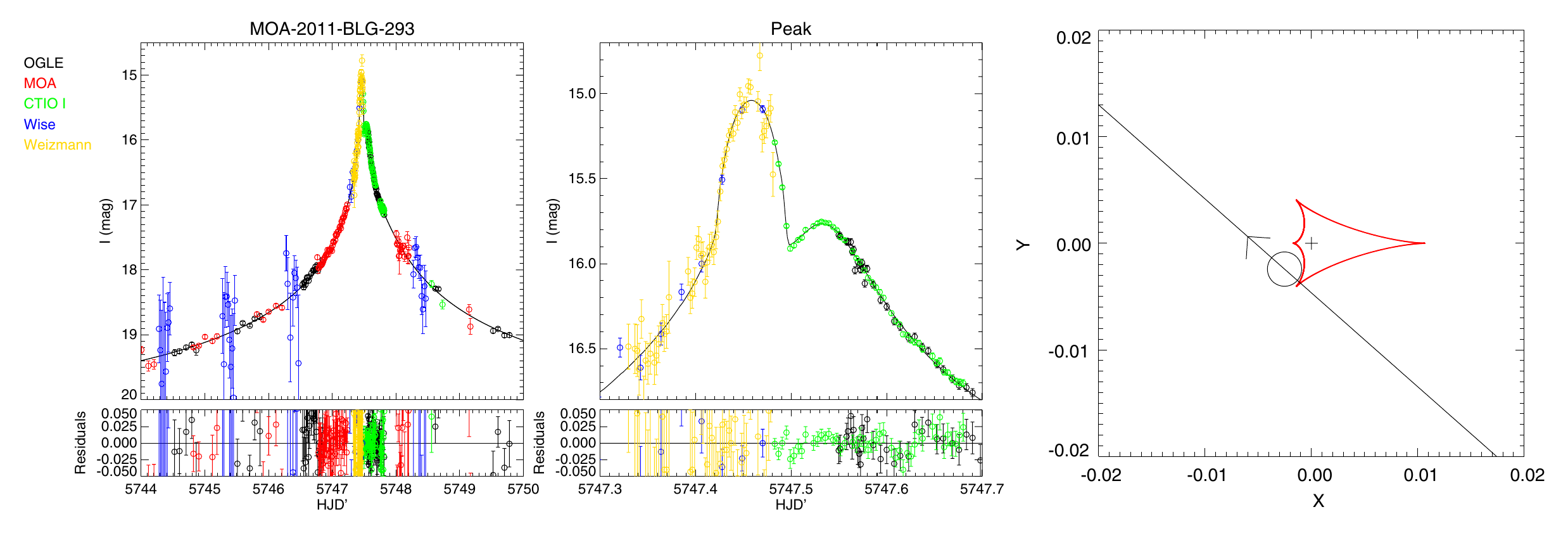}
   }
   \caption{
      \footnotesize{
         Light curve (left and middle panel) and caustic geometry and source trajectory (right panel) for the microlensing event MOA-2011-BLG-293 \citep{yee2012}.
         The deviation from a smooth, temporally symmetric light curve is highlighted by the structure at peak (middle panel) that arises when the source passes over the central caustic (right panel).
         \vspace{1.0mm}
      }
   }
   \label{fig:mb11293_lightcurve}
\end{figure*}

The mass ratio $q$ and the separation $s$ of the two lensing masses define the topology governing the location and morphology of the caustics \citep{erdl1993,dominik1999b}, which are closed curves in the plane of the lens that identify where the magnification of a point-like source formally diverges to infinity.
These caustic curves increase the probability that $\rho$ will be measured, since the finite size of the source will be ``resolved" from the detailed magnification structure of the light curve if the trajectory of the source passes near to or over one or more caustics.
For a lens system comprised of two point masses, there are either one, two, or three non-intersecting caustics.
If the second lensing body is low-mass ($q \ll 1$), there is typically a central caustic located near the primary star and either one (for $s > 1$) or two (for $s < 1$) planetary caustics, whose position and morphology can be approximated analytically for $q \ll 1$ and $s \ne 1$ \citep{bozza2000,chung2005,han2006a}.
For $s \sim 1$, there is one caustic.
See \citet{erdl1993} and \citet{dominik1999b} for the exact values of $s$ where these caustic topologies change for arbitrary $q$.
The light curve, caustic geometry, and source trajectory for an example planetary event are shown in Figure \ref{fig:mb11293_lightcurve}.

%------------------------------------------------------------------------------------------------------------------------
\subsection{From Observables to Parameters} \label{sec:ulens_obs_to_parm}
%------------------------------------------------------------------------------------------------------------------------

%------------------------------------------------------------------------------------------------------------------------
\subsubsection{Observational Methodology} \label{sec:ulens_obs}
%------------------------------------------------------------------------------------------------------------------------

Due to the relatively small detectors that were available at the time when microlensing planet surveys were first initiated, they followed a two-tiered strategy that was first advocated by \citet{gould1992b}.
The microlensing event rate, even toward the Galactic bulge, where the surface density of stars is the highest, is such that an arbitrary source star in the bulge comes within $\sim${\thetaE} of a foreground lensing star only once every $\sim$100,000 years.
To detect a few hundred events per year, it is thus necessary to monitor tens of millions of stars.
Survey telescopes with bigger apertures and the largest available fields of view (FoVs), such as the Optical Gravitational Lensing Experiment (OGLE; \citealt{udalski2003}) and the Microlensing Observations in Astrophysics collaboration (MOA; \citealt{bond2001,sumi2003}), would monitor many tens of square degrees of high stellar density, low extinction fields toward the bulge with cadences of once or twice per night.
These cadences were sufficient to detect and alert the primary events themselves but insufficient to accurately characterize planetary perturbations.
Networks of smaller telescopes, such as the Microlensing Follow-up Network ($\mu$FUN; \citealt{gould2006}) and the Probing Lensing Anomalies NETwork (PLANET; \citealt{albrow1998}), with more readily available narrow-angle detectors, would then monitor a subset of the most promising of these alerted events with the cadence and wider longitudinal coverage required to characterize the planetary anomalies.

Large format detectors, with FoVs of a few square degrees, have facilitated a transition to a phase in which microlensing has been able to increase the planetary yield, by imaging tens of millions of stars in a single pointing with the cadence necessary to detect the primary microlensing events as well as the planet-induced deviations.
Furthermore, surveys performed with these detectors are blind and so circumvent the biases introduced by the reliance on subjectivity and human judgment for the selection of follow-up targets.
Additional groups such as the Wise observatory \citep{shvartzvald2012}, RoboNet \citep{tsapras2009}, and MiNDSTEp \citep{dominik2008,dominik2010} have provided greater access to events through improved longitudinal coverage and higher-cadence observations.
The Korean Microlensing Telescope Network (KMTNet), an array of three 1.6m telescopes located at Cerro Tololo Inter-American Observatory (CTIO) in Chile, South African Astronomical Observatory (SAAO) in South Africa, and Siding Spring Observatory (SSO) in Australia \citep{kim2010,kim2011,kappler2012,poteet2012,atwood2012,kim2016}, represents the next realization of the automated survey strategy, with its ability to conduct a $\sim$16 square-degree survey with a $\sim$10-minute cadence using a homogeneous network of telescopes \citep{sexypants2014a}.

However, for a static two-body lens system, additional information beyond the microlensing observables described in \S \ref{sec:ulens_geometry} is needed to determine the fundamental properties of the planetary system ({\mstar}, {\mpl}, {\dl}).
There are currently two primary methods to achieve this with minimal model dependence, both of which require measuring {\thetaE}, typically by rearranging Equation (\ref{eq:rho}) and combining multiband photometry to determine {\thetastar} with a measurement of $\rho$ through a detection of finite-source effects \citep{yoo2004}.

%------------------------------------------------------------------------------------------------------------------------
\subsubsection{Microlensing Parallax} \label{sec:ulens_parallax}
%------------------------------------------------------------------------------------------------------------------------

The first avenue is by determining {\piE}, which can be accomplished through one or both of two primary channels.
One method involves measuring the distortion in the observed light curve due to the acceleration of the observer relative to the light expected for a constant velocity.
In this situation, the single-platform observer could be the Earth \citep{gould1992a}, a satellite in low-Earth orbit \citep{honma1999}, or a satellite in geosynchronous orbit \citep{gould:2013a}.
This orbital parallax can be measured for events with timescales that are typically a significant fraction of a year and requires good observational coverage (see \citealt{alcock1995,poindexter2005,gaudi2008} for examples).

A second technique involves taking observations from two or more well-separated locations \citep{refsdal1966,hardy1995,gould:1997a}.
Generally this requires two observatories separated by of-order an AU in order to produce detectably different light curves.
This is becoming the dominant mechanism for measuring {\piE} and is referred to as the ``satellite parallax" technique.
It is possible, however, to measure ``terrestrial parallax," which involves two observatories at different longitudes on the Earth monitoring an intrinsically rare high-magnification event, for which ${\uzero} \ll 1$, with extremely high cadence (e.g., \citealt{gould2009,yee2009}).
In all cases, by combining {\piE} with {\thetaE} the total mass of the lensing system can be determined via Equation (\ref{eq:thetaE}), yielding the masses of the individual components of the lensing system.
Furthermore, by assuming the source is located in the Galactic bulge, {\dl} can also be extracted.

\begin{figure}
   \centerline{
      \includegraphics[width=9cm]{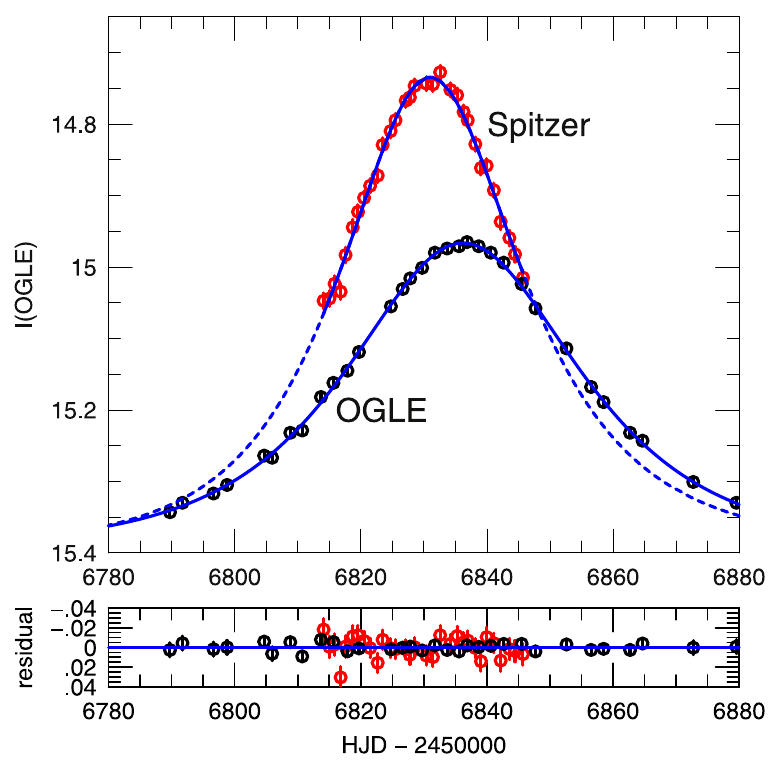}
   }
   \caption{
      \footnotesize{
         Light curve of the microlensing event OGLE-2014-BLG-0939 as seen by {\spitzer} (red points) and OGLE (black points) \citep{yee2015b}.
         The spatial baseline between {\spitzer} and the Earth alters the geometry of the event as seen from each location, inducing a shift in the time and magnitude of the peak amplification of the light from the background source star.
         This shift allows for a measurement of the satellite parallax, helping to determine the mass of and distance to the lensing system, which in this case is an isolated star.
         \vspace{1.0mm}
      }
   }
   \label{fig:ob140939_parallax}
\end{figure}

Recently, {\spitzer} has been employed to measure satellite parallaxes.
\citet{dong2007} first used it to measure the microlensing parallax {\piE} for a weak (i.e., non-caustic-crossing) binary event toward the Small Magellanic Cloud.
A pilot 100-hour program in 2014 made the first satellite parallax measurement of an isolated star \citep{yee2015b}.
The light curve, shown in Figure \ref{fig:ob140939_parallax}, clearly demonstrates the shifts in {\tzero} and {\uzero} that arise from the $\sim$1 AU separation between the Earth and {\spitzer} and that alter the magnification of the source in the ground-based light curve compared to that seen in the space-based light curve.
The 2014 {\spitzer} campaign also resulted in the first satellite parallax measurement for a microlensing exoplanet \citep{udalski2015b}.
In the case of that event, the precision of {\piE} via satellite parallax ($\sim$2.5$\%$) was an order of magnitude better than that obtained through orbital parallax ($\sim$22$\%$), emphasizing the importance of space telescopes for improving the precision on ({\mstar}, {\mpl}, {\dl}).
An 832-hour {\spitzer} campaign in 2015 observed 170 additional events, helped to refine the methodology (see \citealt{yee2015a}), and led to {\piE} measurements for a cold Neptune in the Galactic disk \citep{street2015} and a massive stellar remnant \citep{shvartzvald2015}, among other astrophysically interesting objects.

%------------------------------------------------------------------------------------------------------------------------
\subsubsection{Flux Characterization} \label{sec:ulens_flux}
%------------------------------------------------------------------------------------------------------------------------

The second channel for converting microlensing observables into the fundamental parameters ({\mstar}, {\mpl}, {\dl}) involves constraining the flux of the primary lensing mass: the host star.
Determining {\thetaE} from color information and finite-source effects and assuming a value for {\ds} gives one mass-distance relation for the lens system.
Then, measuring the lens flux {\fl} and applying a mass-luminosity relation \citep{bennett2007} provides a second mass-distance relation, given a value of the extinction toward the lens.
Combining these two allows for the unique determination of ({\mstar}, {\mpl}, {\dl}).
The extinction is known for any line-of-sight within the OGLE-III footprint (see \citealt{nataf2013}).
Therefore, measuring {\fl} gives an additional technique for deriving the fundamental parameters of the lensing system.

It is important to note that this does not necessarily require waiting for the lens and source to be resolved.
In fact, there are several ways by which {\fl} can be constrained, including:
\begin{enumerate}
   \item measuring a color-dependent centroid shift,
   \item imaging the lens after it is spatially resolved from the source,
   \item inferring {\fl} by measuring the elongation of the point spread function (PSF) of the unresolved microlensing target (lens+source) as the lens and source begin to separate, and
   \item promptly obtaining high-resolution follow-up photometry while the lens and source are unresolved.
\end{enumerate}
\citet{drsexypants:2015a} discusses the challenges and possibilities for items 2--4 specifically in the context of KMTNet planetary detections.
\citet{sexypants2014b} furthermore identified the subset of past microlensing events with {\murel} sufficiently high that current high-resolution facilities can spatially resolve the lens and source in $\lesssim$10 years.
Here we focus only on the fourth option.

Measuring {\fl} this way requires near-infrared (NIR) observations at two different epochs: the first while the source is magnified and the event is ongoing, the second with a high-resolution facility after the event is over and the source has returned to its baseline brightness.
By modeling the ground-based light curve, which includes both magnified and unmagnified NIR data, the NIR flux of the source can be measured precisely.
Then, the high-resolution NIR observation at baseline will resolve out all stars not dynamically associated with the event to a high probability.
By subtracting the NIR source flux from the second, unmagnified, observation, any detected flux that is in excess of the source flux can be ascribed to the lens, breaking the degeneracy by searching for the light from the planet's possible host.
We note that this excess light could potentially be due to companions to the lens and/or source instead of, or in addition to, the lens itself.
But, this depends on the underling stellar multiplicity (see, e.g., \citealt{raghavan2010}), and moreover the contamination from undetected, unknown companions to either lens or source is low \citep{drsexypants:2015a}.
The NIR flux characterization method has been applied to a handful of planetary events \citep{bennett2007,dong2009b,janczak2010,sumi2010,batista2011,batista2014,batista2015,fukui2015}.

%%%%%%%%%%%%%%%%%%%%%%%%%%%%%%%%%%%%%%%%%%%%%%%%%%
%%%
\section{Scientific Drivers} \label{sec:science}
%%%
%%%%%%%%%%%%%%%%%%%%%%%%%%%%%%%%%%%%%%%%%%%%%%%%%%

{\ktcn} represents an extraordinary opportunity to make progress in several regimes of exoplanet demographics.
Covering 3.7 deg$^{2}$, it will be the first space-based blind survey dedicated to exoplanetary microlensing, facilitating {\piE} measurements for $\gtrsim$127 events (see \S \ref{sec:k2_pixels}).
In contrast with the 2014 and 2015 {\spitzer} programs, which require $\gtrsim$4 days between target selection and observation, {\ktcn} will be able to measure {\piE} for short-timescale events ({\tE} of-order 1 day), which are potentially indicative of free-floating planets (FFPs).
Microlensing's intrinsic sensitivity to bound planets beyond the snow line makes it an indispensable complement to radial velocity, transit, and direct imaging exoplanet searches that seek to improve demographic understanding and provide input for planet formation models.
Furthermore, planetary systems with satellite parallax constraints will better our understanding of the frequency and distribution of planets at a wide range of distances from Earth.
We lastly note that, as with {\spitzer}, it will be possible to probe the stellar remnant population \citep{shvartzvald2015}, measure the mass of isolated objects such as stars and brown dwarfs \citep{zhu2015a}, and determine the fundamental parameters for binary star systems \citep{zhu2015b}.

%------------------------------------------------------------------------------------------------------------------------
\subsection{Free-floating Planets} \label{sec:science_ffps}
%------------------------------------------------------------------------------------------------------------------------

\begin{figure}
   \centerline{
      \includegraphics[width=9cm]{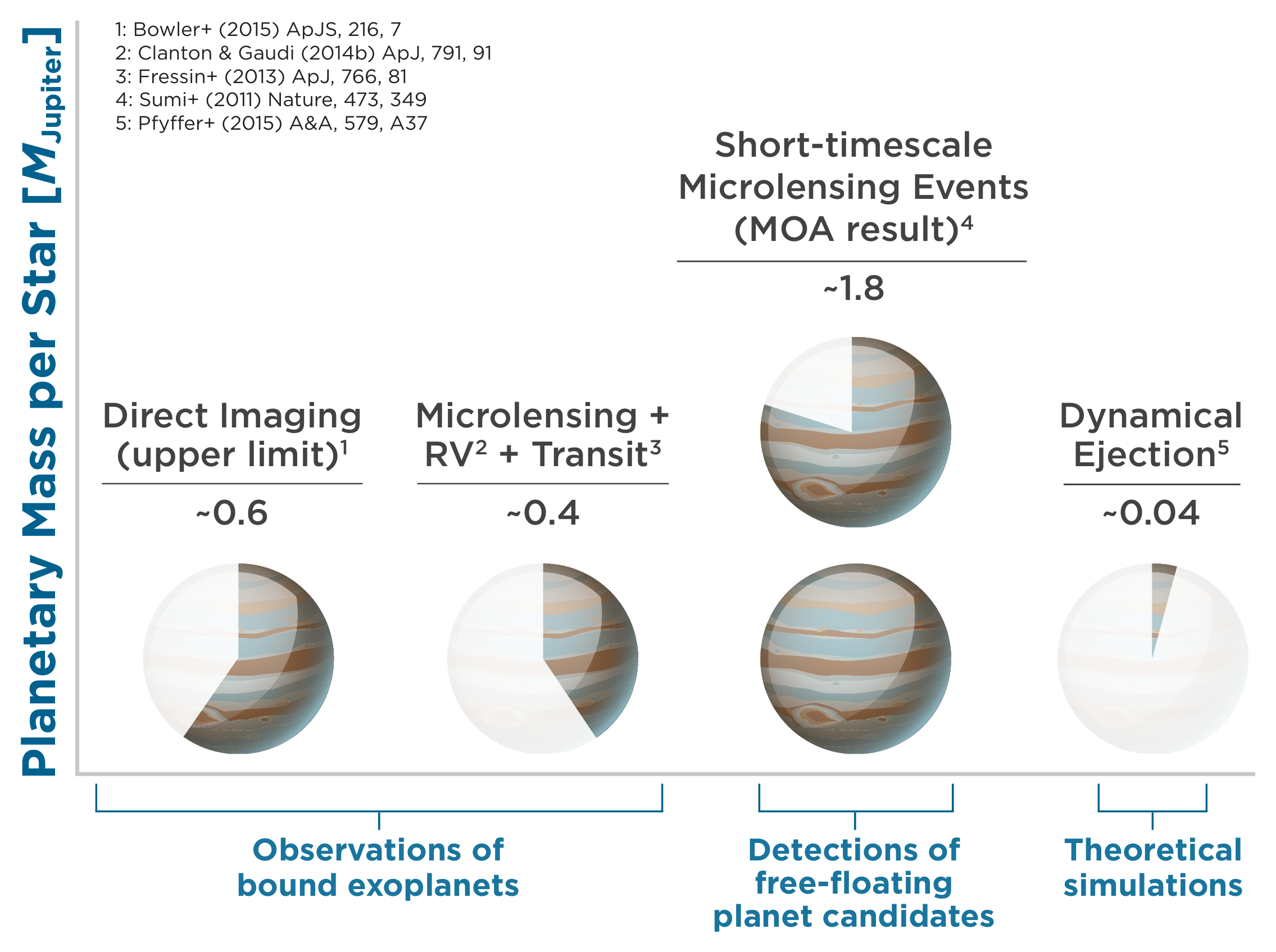}
   }
   \caption{
      \footnotesize{
         Estimates of planetary-mass material per star from different observational techniques and for theoretical predictions.
         MOA's result indicates that free-floating planet (FFP) candidates account for $\sim$1.8 $M_{\rm Jup}$ per star \citep{sumi2011}.
         The upper limit of planetary material bound to stars from direct imaging \citep{bowler2015} yields, at most, one-third of this amount.
         Including transit results from {\kepler} \citep{fressin2013} and RV and microlensing planets \citep{clanton2014b} brings the total mass of bound planets per star to $\sim$1.0 $M_{\rm Jup}$.
         Thus, if all FFP candidates are truly FFPs, these objects dominate the mass budget of planet formation.
         Moreover, simulations of gravitational dynamics during planetary formation and evolution do not predict the number of FFPs predicted by MOA.
         For example, the most optimistic simulations by \citet{pfyffer2015}, which do not include eccentricity or inclination damping, produce only $\sim$0.04 $M_{\rm Jup}$ of ejected planets per star.
         Deriving the true mass function of FFPs with {\ktcn} will address and help resolve this tension.
         We note that the accounting presented here is extremely rough and is intended to give an idea of the scale of the problem rather than a precise quantitative description.
         \vspace{1.0mm}
      }
   }
   \label{fig:mass_comparison}
\end{figure}

\citet{sumi2011} announced the discovery of an excess of short-timescale microlensing events, with $t_{\rm E} < 2$ days, discovered by the MOA survey, which they inferred to be caused by a population of ``unbound or distant planetary-mass" objects with masses comparable to that of Jupiter and outnumbering main sequence stars by 2:1.
Their results imply that these FFP candidates account for $\sim$1.8 $M_{\rm Jup}$ of planetary-mass objects per star on the main sequence, as highlighted in Figure \ref{fig:mass_comparison}.

Such a plenitude of FFPs stands in stark contrast to observational constraints on bound planetary systems as well as theoretical expectations for ejected planets.
Combining the detailed statistical analysis of exoplanets discovered by microlensing and radial velocity surveys out to an orbital period of $\sim$10$^{5}$ d \citep{clanton2014b} and including planets with small radii inaccessible to RV surveys and planets orbiting more massive host stars \citep{fressin2013} only accounts for $\sim$0.4 $M_{\rm Jup}$ of bound planetary mass material per star.
Extending out to the farthest reaches of stellar systems and adding, optimistically, cold-start-based upper limits from direct imaging searches for loosely bound planets around young stars only allows for, at most, an additional $\sim$0.6 $M_{\rm Jup}$ \citep{bowler2015}.
Furthermore, current theories of planetary dynamics cannot explain the existence of such an abundance of FFPs.
For example, simulations by \citet{pfyffer2015} of the formation and evolution of planetary systems without eccentricity damping eject only $\sim$0.04 $M_{\rm Jup}$ of planets per star, a rate that is significantly lower than is needed to explain the MOA result.
We note that the accounting presented here is extremely rough and is intended to give an idea of the scale of the problem rather than a precise quantitative description.

If the short-timescale events discovered by MOA are in fact FFPs, these objects must thus dominate the mass budget of planet formation.
Additionally, their abundance is severely underestimated by even the most detailed theoretical models of planetary dynamics.
However, short-timescale microlensing events can also be caused by stars with large proper motions in the Galactic bulge or low-mass planets that are bound to but widely separated from their host star.
It is thus of crucial importance to investigate the nature of events with short $t_{\rm E}$ and determine whether they are indeed caused by free-floating planetary-mass objects.

Satellite parallax measurements made with {\kt} during {\cn} will help verify whether the cause of each of these short-timescale events is, in fact, a low-mass object.
The NIR source flux measurements enabled with ground-based facilities (see \S \ref{sec:ground}) will then set the stage for follow-up high-resolution NIR observations that will help distinguish between a planet that is bound to but widely separated from its host star and one that is truly free-floating (see \S \ref{sec:ulens_flux}).

{\ktcn} presents another method for vetting FFP candidates.
{\kt} will take continuous observations with a photometric precision that may be better than that attained by many ground-based telescopes.
There will thus be two source trajectories, one as seen \textit{continuously} by {\kepler} and one as seen from the Earth.
Together they increase the geometric probability of detecting potential host stars, and during {\ktcn} this will be done with an efficiency that is much higher than was possible for the \citet{sumi2011} sample.

%------------------------------------------------------------------------------------------------------------------------
\subsection{Galactic Distribution of Exoplanets} \label{sec:science_galactic_dist}
%------------------------------------------------------------------------------------------------------------------------

\begin{figure}
   \centerline{
      \includegraphics[width=9cm]{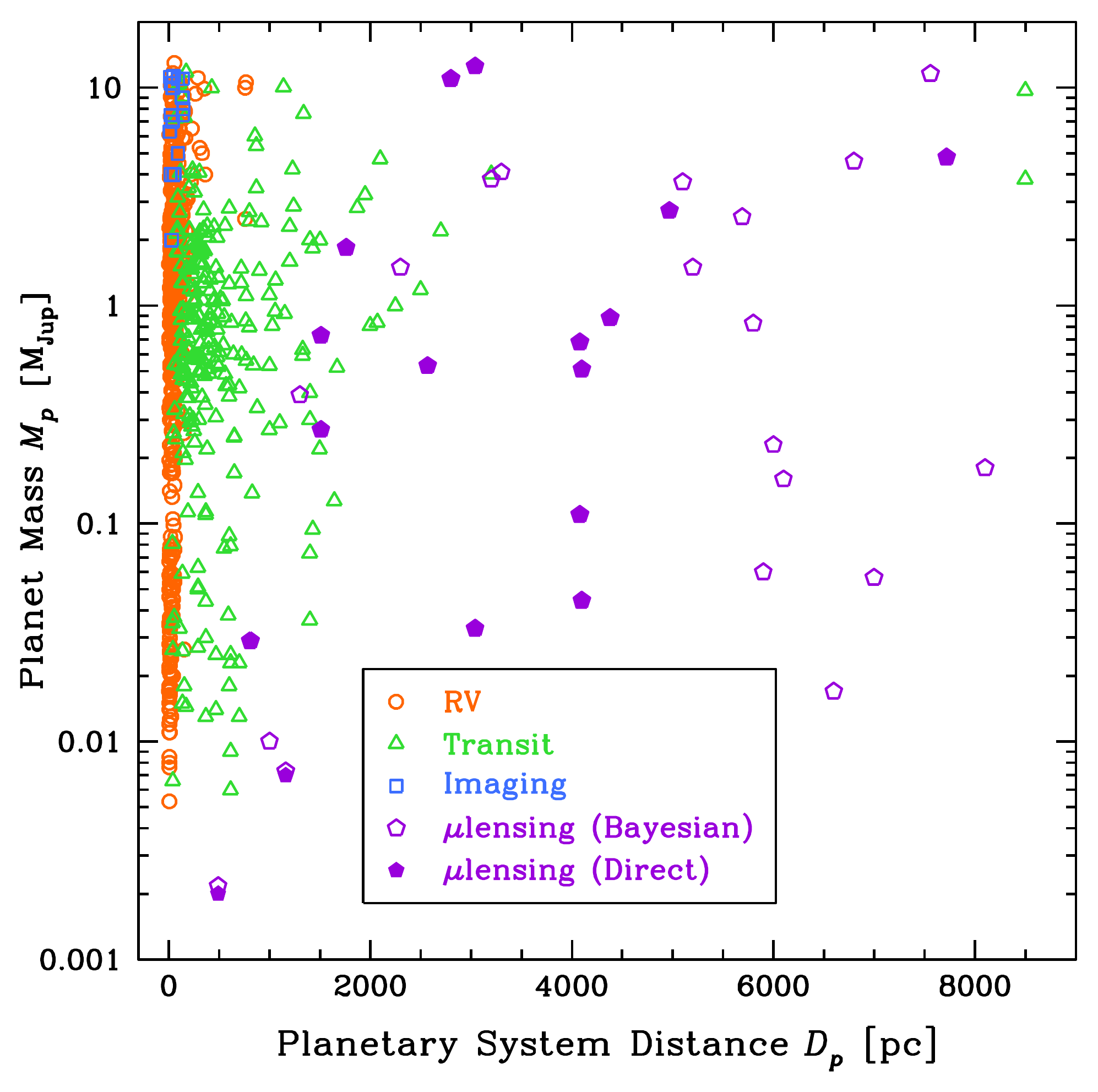}
   }
   \caption{
      \footnotesize{
         Planet mass {\mpl} as a function of the distance to the planetary system from Earth {\dpl}, with points styled according to discovery technique.
         The data were taken from the NASA Exoplanet Archive \citep{akeson2013}.
         Of the 844 planets shown here, only 34 were discovered by microlensing.
         However, microlensing exoplanets are responsible for the vast majority of known systems with ${\dpl} \gtrsim 2000$ pc ($\sim$70$\%$).
         Efforts such as {\ktcn} will improve our understanding of planet demographics throughout the Galaxy by helping to directly measure planet distances out to the bulge.
         \vspace{1.0mm}
      }
   }
   \label{fig:planet_distance}
\end{figure}

Figure \ref{fig:planet_distance} shows planet mass {\mpl} as a function of planetary system distance from Earth {\dpl} for all verified exoplanets, with {\mpl} and {\dpl} data taken from the NASA Exoplanet Archive\footnote{\url{http://exoplanetarchive.ipac.caltech.edu/}} \citep{akeson2013}.
While the 34 microlensing detections account for only $\sim$4$\%$ of the 844 total such planets, they constitute $\sim$40$\%$ (32 out of 78) of those with ${\dpl} > 1000$ pc and $\sim$70$\%$ (26 out of 36) of those with ${\dpl} > 2000$ pc.
We note that only half of the microlensing-discovered exoplanets have mass constraints either via {\piE} or NIR flux measurements, while the rest rely on characterization through Bayesian analysis.
There are thus only 17 microlensing planets (in 15 systems) with directly measured distances.

In order to best understand the frequency of planets in different stellar environments, it is crucial that any selection effects be well understood (see \citealt{street2015}), a problem made tractable by the {\ktcn} automated survey.
Such an approach will not only then improve our understanding of planet demographics from the Solar neighborhood to the Galactic bulge but will also allow us to investigate planet frequency in the Galactic disk versus the Galactic bulge \citep{calchinovati2015,penny2016}, or the occurrence rate as a function of, e.g., metallicity \citep{montet2014}.
Perhaps most compelling is that bound planetary systems with satellite parallax-derived masses and distances are invaluable as we strive toward a comprehensive picture of exoplanet demographics that can reconcile detections obtained using multiple techniques \citep{clanton2014a,clanton2014b,clanton2015}.

%%%%%%%%%%%%%%%%%%%%%%%%%%%%%%%%%%%%%%%%%%%%%%%%%%
%%%
\section{{\ktcn} Observational Setup} \label{sec:k2_parms}
%%%
%%%%%%%%%%%%%%%%%%%%%%%%%%%%%%%%%%%%%%%%%%%%%%%%%%

\citet{gould2013b} identified that a repurposed {\kepler} spacecraft could be utilized as a microlens parallax satellite.
They estimated that a 90-day survey of the Ecliptic that is coordinated with ground-based observatories would result in {\piE} measurements for several hundred microlensing events, including $\sim$12 planetary in nature.
{\ktcn} is, in essence, a realization of this idea.

%------------------------------------------------------------------------------------------------------------------------
\subsection{Campaign and Spacecraft Parameters} \label{sec:k2_campaign_spacecraft}
%------------------------------------------------------------------------------------------------------------------------

\begin{figure*}
   \centerline{
      \includegraphics[width=18cm]{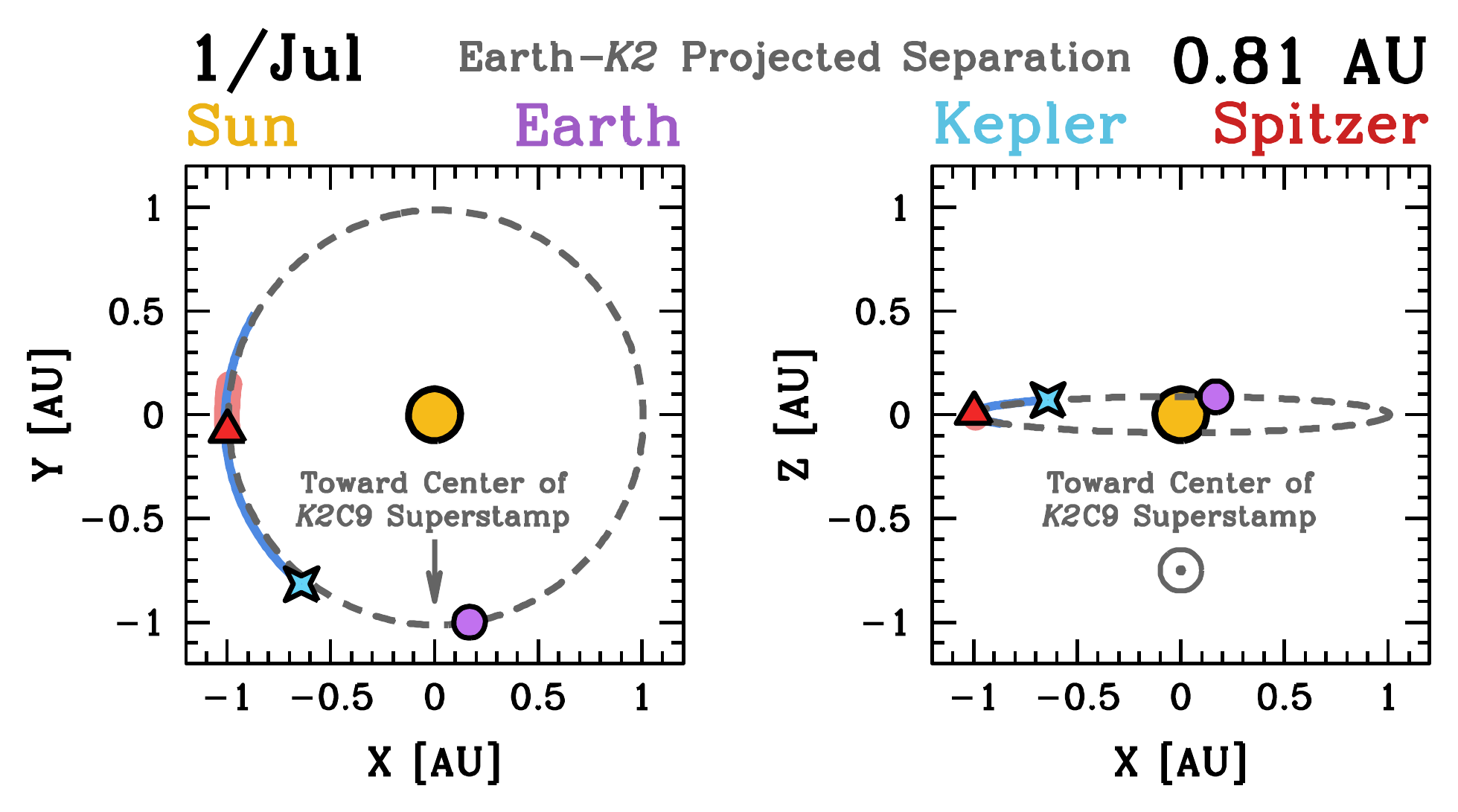}
   }
   \caption{
      \footnotesize{
         The temporal evolution of the orbits of the Earth, {\kepler}, and {\spitzer} throughout {\cn} as seen from the center of the {\ktcn} superstamp.
         In the right panel, the tilt of the Ecliptic arises from the fact that the approximate center of the superstamp is $\sim$5 degrees below the Ecliptic (see Figure \ref{fig:k2c9_superstamp}).
         Thus, in the right panel, the directional label indicates that the line-of-sight toward the center of the superstamp extends out of the page (and does not mark the (X,Z) coordinate for the superstamp).
         We have furthermore created short videos to help visualize and make intuitive the satellite parallax effect as it will be measured during {\ktcn}.
         Both still-frame pdfs and animated gifs, along with a brief README file, can be found here: \url{https://osu.app.box.com/k2c9animations}.
         \vspace{1.0mm}
      }
   }
   \label{fig:earth_k2_spitzer_orbits}
\end{figure*}

{\ktcn} will conduct an 86-day microlensing survey toward the Galactic bulge from 7/April through 1/July of 2016.
The spacecraft will be re-oriented to point along its velocity vector (+VV), enabling it to observe the bulge during a window when it is simultaneously visible for ground-based telescopes.
The field center for {\cn} is located at (RA, Dec) = (18:01:25, -21:46:47).
A minimum of 2.8 million pixels, or 3.4 deg$^{2}$, will be dedicated to the microlensing survey, with the remaining $\sim$15$\%$ of the downlinkable area devoted to the {\kt}'s Director's Discretionary Target program.
In \S \ref{sec:k2_pixels} we discuss the methodology used to determine the exact superstamp, or roughly contiguous selection of pixels to be downlinked, that will comprise the microlensing survey area for {\ktcn}.

Figure \ref{fig:earth_k2_spitzer_orbits} shows the orbits of the Earth, {\kepler}, and {\spitzer} (which will contribute simultaneous observations for the final 13 days of {\cn}; see \S \ref{sec:spitzer}) throughout {\cn}.
The projected separation between {\kepler} and the Earth as viewed from the center of the {\ktcn} superstamp, $D_{\perp}$, changes throughout the duration of the campaign, and dictates the range of {\thetaE} for which the geometry will be most favorable for measuring {\piE}.
We have created short movies to help visualize the temporal evolution of $D_{\perp}$ over the course of {\cn} and to facilitate intuition about the satellite parallax effect.
To browse and utilize both still-frame pdfs and animated gifs, please visit:

\vspace{0.5mm}
\begin{center}
   \url{https://osu.app.box.com/k2c9animations}
\end{center}
\vspace{0.5mm}

In addition to the re-orientation of the spacecraft, {\ktcn} will feature several modifications to its standard observing procedure.
To attain the survey area quoted above, {\ktcn} will utilize a mid-campaign data downlink in order to increase the number of microlensing events for which {\piE} will be measured.
This will divide the campaign into two halves denoted as {\cn}a and {\cn}b.
Furthermore, careful exploration by the {\kt} team has approved the possibility to add, to the target list for both {\cn}a and {\cn}b, postage stamps for individual microlensing events that will have been detected by the ground-based survey groups (see \S \ref{sec:ground_surveys}) and that will be ongoing and expected to peak during {\ktcn}, increasing the number of events for which {\piE} can be measured.
The deadlines for including ongoing events are listed in Table \ref{tab:k2c9_parms}.
A postage stamp for such an ongoing event will consist of a square of a few hundred pixels.
To account for the additional data to be downlinked, each campaign half will be shortened by an amount of time that is proportional to the number of postage stamps included in the target list for these ongoing events.
Given the low fractional cost of such events, the time required to account for these ongoing events will be small: of-order 20 minutes per 1,000 pixels, a factor that is likely less than the uncertainty in the data storage requirements onboard the spacecraft.
We discuss the implementation and projected yields of this endeavor in \S \ref{sec:k2_pixels}.

All observations will be long cadence (i.e., 30-minute sampling).
Within the first week after {\cn}a and {\cn}b have each concluded, the corresponding cadence pixel files will be available through the Mikulski Archive for Space Telescopes (MAST)\footnote{\url{https://archive.stsci.edu/}}.
The pipeline-processed target pixel files for the full campaign will be posted to MAST on 26/September/2016.
Table \ref{tab:k2c9_parms} describes all of the observational parameters for {\ktcn}.

\begin{deluxetable}{lcr}
\centering
\tablecaption{{\ktcn} Observational Parameters}
\tablewidth{0pt}
\tablehead{
}
\startdata
\textbf{Key Dates$^{a}$}   &      &      \\
\hspace{5.0mm} \textit{Initial superstamp selection deadline}   &      &   25/January   \\
\hspace{5.0mm} \textit{Augmented superstamp submission}   &      &   18/February   \\
\hspace{5.0mm} \textit{{\cn}a}   &      &      \\
\hspace{10.0mm} Ongoing event upload deadline   &      &   1/March   \\
\hspace{10.0mm} Observing window   &      &   7/April--18/May   \\
\hspace{10.0mm} Raw data available at MAST   &      &   24/May   \\
\hspace{5.0mm} \textit{Mid-campaign break (data downlink)}   &      &   19--21/May   \\
\hspace{5.0mm} \textit{{\cn}b}   &      &      \\
\hspace{10.0mm} Ongoing event upload deadline   &      &   25/April   \\
\hspace{10.0mm} Observing window   &      &   22/May--1/July   \\
\hspace{10.0mm} Raw data available at MAST   &      &   6/July   \\
\hspace{5.0mm} \textit{Processed data available at MAST}   &      &   26/September   \\
\textbf{Superstamp Center (approximate)}   &      &      \\
\hspace{5.0mm} RA (hh:mm)   &      &   17:57   \\
\hspace{5.0mm} Dec (dd:mm)   &      &   -28:24   \\
\textbf{Aperture} [m]   &      &   0.95   \\
\textbf{Plate Scale} [$\arcsec$ pixel$^{-1}$]   &      &   3.98   \\
\textbf{Pixels$^{b}$} [$\times$10$^{6}$]   &      &   3.06   \\
\textbf{Survey Area$^{b}$} [deg$^{2}$]   &      &   3.7   \\
\textbf{Cadence} [min]   &      &   30
\enddata
\tablenotetext{a}{All dates are in 2016.}
\tablenotetext{b}{This refers to the area of the final, augmented superstamp pixel selection.}
\label{tab:k2c9_parms}
\vspace{1.0mm}
\end{deluxetable}

%------------------------------------------------------------------------------------------------------------------------
\subsection{Pixel Selection} \label{sec:k2_pixels}
%------------------------------------------------------------------------------------------------------------------------

The {\kt} camera has a full FoV of 105 deg$^{2}$.
However, only a few percent of the pixels can be downlinked due to limited data storage.
Prior {\kt} campaigns thus observed a postage stamp of pixels for each individual pre-selected target star.
Since {\cn} will conduct an automated survey to detect lensing events, which are transient and inherently unpredictable, the pixels that will be downlinked will instead form a roughly contiguous region, or superstamp.
The highest scientific return of {\ktcn} comes from the events that are observed both from {\kt} and from the ground.
We select the survey superstamp to optimize the predicted number of events observed from Earth since the ground-based event rate is far better understood than that expected for {\kt}.

To predict the ground-based event rate across the full {\ktcn} FoV we use the framework presented by \citet{poleski2016}.
He showed that the number of standard events, or events well-described by a single-lens model (i.e., excluding two-body lensing events or single-lens events with strong finite-source effects), detected by the OGLE-III survey is a linear function of the product of two observables that can be measured relatively easily: the surface density of red clump (RC) stars ($N_{\rm RC}$), and the surface density of all stars brighter than the completeness limit ($N_*(I<20~{\rm mag})$).
The reasoning behind this linear relation is a simple model: the event rate should be a product of number of potential lenses (which correlates with $N_{\rm RC}$) and the number of potential sources (approximated by $N_*(I<20~{\rm mag})$).
RC stars are used because it is possible to use a color-magnitude diagram to identify and count them in all but the highest-extinction regions.
The final formula of \citet{poleski2016} modifies this product slightly by varying the brightness limit and the exponent of $N_{\rm RC}$ as such:
%%% Look here for how to split equations that are too long: http://moser-isi.ethz.ch/docs/typeset_equations.pdf
\begin{IEEEeqnarray}{rCl} \label{equ:p15fin}
   \frac{\gamma(t_{\rm E} > 8~{\rm d})}{{\rm deg^{-2} yr^{-1}}} & = & 0.767 \left(\frac{N_{\rm RC}}{10^3~{\rm deg^{-2}}}\right)^{0.55} \nonumber\\ && \times \>
   \left(\frac{N_{*}(I < 20.5~{\rm mag})}{10^6~{\rm deg^{-2}}}\right)  - 14.6,
\end{IEEEeqnarray}
to better fit the data.
Here $\gamma(t_{\rm E} > 8~{\rm d})$ is the observed number of standard events per year per ${\rm deg^2}$ with {\tE} longer than $8~{\rm d}$.
The limit on the event timescale was applied in order to reduce the impact of the varying OGLE-III observing cadence, and it removed $14\%$ of events in the best-observed fields.
The $\gamma(t_{\rm E} > 8~{\rm d})$ was estimated based on the catalog of standard microlensing events in the OGLE-III survey \citep{wyrzykowski2015}.
The OGLE-III catalog is the largest database of microlensing events selected in a uniform way and with minimal contamination from false positives.
\citet{poleski2016} limited the sample to events observed in fields with an average of 165 epochs per year.
The $N_{\rm RC}$ values were taken from \citet{nataf2013} and $N_*$ values were calculated based on \citet{szymanski2011}.

The {\ktcn} superstamp should balance yielding the highest possible event rate with facilitating ground-based tiling strategies with the highest cadences and coverage of the superstamp.
Such a task is complicated by the fact that, unlike the {\ktcn} footprint, most ground-based cameras are aligned with the equatorial coordinate grid.
For the initial superstamp pixel selection we divide every {\kt} channel of $1100\times1024$ pixels into an $11\times10$ grid, resulting in $6.6\arcmin \times 6.8\arcmin$ regions.
Each such region is included or excluded as a whole.
Some regions that have a high expected event rate will subtend small areas with high extinction that do not contribute to the event rate, but we do not exclude these sub-regions.
Some of the bulge regions with high event rate are beyond the OGLE-III footprint and thus are not included in the RC density study by \citet{nataf2013}.
However, we are able to extrapolate the \citet{nataf2013} $N_{\rm RC}$ values since they correlate with the Galactic bulge density profile of \citet{kent1992}.

We use this correlation to estimate $N_{\rm RC}$ across the entire {\ktcn} footprint.
We find $N_*(I<20.5~{\rm mag})$ values using the reference images of the ongoing OGLE-IV survey \citep{udalski2015a}.
The OGLE-IV reference images do not cover the full Galactic bulge, but the missing areas show low event rates in optical bands.
We estimate the event rate for each $6.6\arcmin \times 6.8\arcmin$ region and select those with the highest event rate until we have accumulated the initial allocation for the total survey area of 2.8 million pixels.
One of the regions with the highest event rate as selected in this manner is located in the northern bulge region at $(l, b) = (2.8^{\circ}, +3.7^{\circ})$.
Given the high observational cost of covering this single $6.6\arcmin \times 6.8\arcmin$ region from the ground, we reject it from the final {\ktcn} superstamp.
All other selected regions fall in five {\kt} channels.
The initial superstamp selection was made public and the observing strategy for some of the ground-based resources was determined so as to have the maximum overlap with this selection. 

Close to the pixel selection deadline it turned out that the {\kt} Director's Discretionary Target program used fewer pixels than were initially allocated to it.
As a result, the remaining pixel resources were devoted to the microlensing experiment, increasing the area for the {\cn} microlensing survey.
We select the additional $6.6\arcmin \times 6.8\arcmin$ regions only in the five previously chosen channels. 
Finally, we modify the shape of the superstamp on a smaller scale in order to have the best overlap with large ground-based observing programs, the footprints were known by that time. 
We try to keep the shape of the superstamp relatively simple and continuous inside each channel. 

\begin{figure*}
   \centerline{
      \includegraphics[width=18cm]{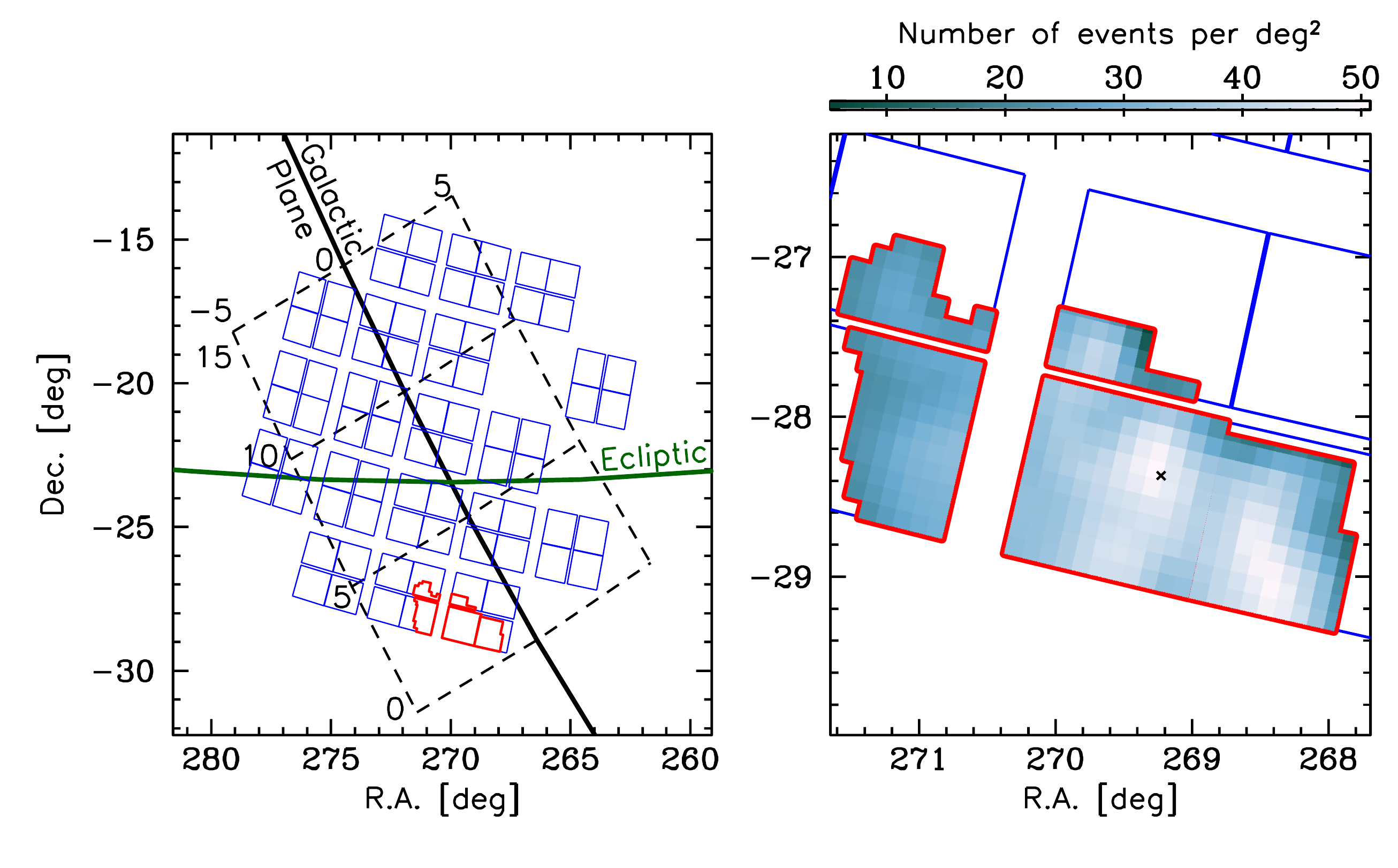}
   }
   \caption{
      \footnotesize{
         The full {\ktcn} FoV (outlined in blue) and the final superstamp (red) as selected by the methodology described in \S \ref{sec:k2_pixels}.
         In the right panel the black cross identifies the center of the region used to create Figure \ref{fig:stardensity_cdf}.
         These 3.06 million pixels, or 3.7 deg$^{2}$ (see Table \ref{tab:k2c9_parms} for all {\ktcn} parameters), will produce an estimated $\sim$127 events that will peak in the ground-based data during the campaign.
         Some number of ongoing events located outside the superstamp may be added to the target lists for {\cn}a and {\cn}b.
         \vspace{1.0mm}
      }
   }
   \label{fig:k2c9_superstamp}
\end{figure*}

Figure \ref{fig:k2c9_superstamp} shows the 3.06 million pixels, or 3.7 deg$^{2}$ (see Table \ref{tab:k2c9_parms} for all {\ktcn} parameters), that comprise the final {\ktcn} superstamp.
Equation (\ref{equ:p15fin}) predicts that 110 standard events will occur within the superstamp throughout the entire bulge observing season (early February through early November).
Including events with $t_{\rm E} > 8~{\rm d}$ and scaling to the higher cadence of OGLE-IV ($20~{\rm min}$) gives 300 events.
Including non-standard events results in as many as 337 events.
Out of these, 127 should peak during {\ktcn}, but it is not guaranteed that the peak for every event will be seen in the {\kt} data, particularly given the shift induced by the satellite parallax.
Similarly, there can be events that are found in the {\kt} data that are below the detection threshold for ground-based surveys.
We note that in some cases the microlens parallax can be measured using ground-based and satellite photometry event if a satellite did not observe the peak of the event \citep{calchinovati2015}.
Additional events will probably be recovered in the ground-based data after the campaign.

The final superstamp was sent to NASA on 18/February/2016.
However, even after this date it will be possible to add to the target list postage stamps that correspond to microlensing events within the {\ktcn} FoV (but outside the superstamp) that have been detected by the ground-based surveys.
The deadline for adding events that are expected to peak at any point during {\ktcn} is 1/March/2016.
In total, 34 ongoing events were added to the target list for {\cn}a.
Table \ref{tab:c9a_ongoing_events} provides a list of the event names, coordinates, and baseline $I$-band magnitudes.
Similarly, the deadline for events detected by ground-based surveys that are expected to peak during {\cn}b is 25/April/2016.
Given the morphology of the microlensing event rate across the bulge, the majority of these events will be located close to the survey superstamp.
Furthermore, only a subset of these additional ongoing events will actually \textit{peak} during {\ktcn} because of the relatively long delay between the selection dates and the start of observations.

\begin{deluxetable*}{lccc}
\tablecaption{Ongoing Events Added to {\ktcn}a Target List$^{a}$}
\tablewidth{0pt}
\tablehead{
\multicolumn{1}{c}{Event$^{b,c,d}$}                     &
\multicolumn{1}{c}{R.A. (J2000)}                        &
\multicolumn{1}{c}{Dec. (J2000)}                        &
\multicolumn{1}{c}{Baseline $I$-band magnitude$^{e}$}
}
\startdata
OGLE-2016-BLG-0122                      &   17$^{\rm h}$37$^{\rm m}$48$^{\rm s}$\hskip-2pt.66   &   -26$^{\rm h}$24$^{\rm m}$50$^{\rm s}$\hskip-2pt.5   &   17.9   \\
OGLE-2016-BLG-0037/OGLE-2016-BLG-0095   &   17$^{\rm h}$41$^{\rm m}$52$^{\rm s}$\hskip-2pt.14   &   -25$^{\rm h}$53$^{\rm m}$52$^{\rm s}$\hskip-2pt.9   &   19.5   \\
OGLE-2016-BLG-0144                      &   17$^{\rm h}$41$^{\rm m}$29$^{\rm s}$\hskip-2pt.18   &   -26$^{\rm h}$17$^{\rm m}$36$^{\rm s}$\hskip-2pt.7   &   20.2   \\
OGLE-2016-BLG-0119                      &   17$^{\rm h}$43$^{\rm m}$57$^{\rm s}$\hskip-2pt.28   &   -21$^{\rm h}$27$^{\rm m}$50$^{\rm s}$\hskip-2pt.2   &   19.4   \\
OGLE-2016-BLG-0129                      &   17$^{\rm h}$42$^{\rm m}$57$^{\rm s}$\hskip-2pt.98   &   -24$^{\rm h}$33$^{\rm m}$39$^{\rm s}$\hskip-2pt.5   &   19.7   \\
OGLE-2016-BLG-0041                      &   17$^{\rm h}$43$^{\rm m}$07$^{\rm s}$\hskip-2pt.82   &   -26$^{\rm h}$47$^{\rm m}$12$^{\rm s}$\hskip-2pt.7   &   17.7   \\
OGLE-2016-BLG-0022                      &   17$^{\rm h}$45$^{\rm m}$01$^{\rm s}$\hskip-2pt.50   &   -22$^{\rm h}$15$^{\rm m}$43$^{\rm s}$\hskip-2pt.1   &   19.3   \\
OGLE-2016-BLG-0065                      &   17$^{\rm h}$46$^{\rm m}$42$^{\rm s}$\hskip-2pt.65   &   -23$^{\rm h}$24$^{\rm m}$38$^{\rm s}$\hskip-2pt.1   &   18.0   \\
OGLE-2016-BLG-0021                      &   17$^{\rm h}$47$^{\rm m}$23$^{\rm s}$\hskip-2pt.67   &   -22$^{\rm h}$18$^{\rm m}$57$^{\rm s}$\hskip-2pt.2   &   20.8   \\
OGLE-2016-BLG-0091                      &   17$^{\rm h}$47$^{\rm m}$53$^{\rm s}$\hskip-2pt.69   &   -24$^{\rm h}$14$^{\rm m}$20$^{\rm s}$\hskip-2pt.1   &   17.4   \\
OGLE-2016-BLG-0068                      &   17$^{\rm h}$50$^{\rm m}$39$^{\rm s}$\hskip-2pt.12   &   -22$^{\rm h}$18$^{\rm m}$30$^{\rm s}$\hskip-2pt.0   &   20.4   \\
OGLE-2016-BLG-0052                      &   17$^{\rm h}$51$^{\rm m}$21$^{\rm s}$\hskip-2pt.10   &   -23$^{\rm h}$18$^{\rm m}$38$^{\rm s}$\hskip-2pt.4   &   18.8   \\
OGLE-2016-BLG-0066                      &   17$^{\rm h}$52$^{\rm m}$56$^{\rm s}$\hskip-2pt.62   &   -21$^{\rm h}$45$^{\rm m}$48$^{\rm s}$\hskip-2pt.6   &   21.8   \\
OGLE-2016-BLG-0053                      &   17$^{\rm h}$52$^{\rm m}$30$^{\rm s}$\hskip-2pt.73   &   -22$^{\rm h}$38$^{\rm m}$53$^{\rm s}$\hskip-2pt.7   &   18.7   \\
OGLE-2016-BLG-0056                      &   17$^{\rm h}$57$^{\rm m}$01$^{\rm s}$\hskip-2pt.01   &   -21$^{\rm h}$15$^{\rm m}$56$^{\rm s}$\hskip-2pt.4   &   18.0   \\
OGLE-2016-BLG-0027                      &   18$^{\rm h}$03$^{\rm m}$47$^{\rm s}$\hskip-2pt.67   &   -26$^{\rm h}$36$^{\rm m}$05$^{\rm s}$\hskip-2pt.2   &   19.4   \\
MOA-2016-BLG-023/OGLE-2016-BLG-0079     &   18$^{\rm h}$06$^{\rm m}$52$^{\rm s}$\hskip-2pt.43   &   -26$^{\rm h}$44$^{\rm m}$01$^{\rm s}$\hskip-2pt.5   &   16.0   \\
OGLE-2016-BLG-0083                      &   18$^{\rm h}$06$^{\rm m}$27$^{\rm s}$\hskip-2pt.61   &   -27$^{\rm h}$53$^{\rm m}$32$^{\rm s}$\hskip-2pt.9   &   17.5   \\
OGLE-2016-BLG-0127                      &   18$^{\rm h}$06$^{\rm m}$08$^{\rm s}$\hskip-2pt.61   &   -26$^{\rm h}$33$^{\rm m}$38$^{\rm s}$\hskip-2pt.1   &   20.0   \\
OGLE-2016-BLG-0211                      &   18$^{\rm h}$07$^{\rm m}$14$^{\rm s}$\hskip-2pt.08   &   -27$^{\rm h}$50$^{\rm m}$31$^{\rm s}$\hskip-2pt.1   &   18.3   \\
OGLE-2016-BLG-0078                      &   18$^{\rm h}$08$^{\rm m}$20$^{\rm s}$\hskip-2pt.13   &   -26$^{\rm h}$57$^{\rm m}$40$^{\rm s}$\hskip-2pt.0   &   18.6   \\
OGLE-2015-BLG-2112                      &   18$^{\rm h}$08$^{\rm m}$12$^{\rm s}$\hskip-2pt.16   &   -24$^{\rm h}$56$^{\rm m}$47$^{\rm s}$\hskip-2pt.0   &   15.8   \\
OGLE-2016-BLG-0082                      &   18$^{\rm h}$08$^{\rm m}$15$^{\rm s}$\hskip-2pt.43   &   -27$^{\rm h}$51$^{\rm m}$24$^{\rm s}$\hskip-2pt.4   &   16.1   \\
OGLE-2016-BLG-0077/MOA-2016-BLG-052     &   18$^{\rm h}$08$^{\rm m}$25$^{\rm s}$\hskip-2pt.76   &   -27$^{\rm h}$13$^{\rm m}$01$^{\rm s}$\hskip-2pt.7   &   17.4   \\
OGLE-2016-BLG-0230                      &   18$^{\rm h}$13$^{\rm m}$15$^{\rm s}$\hskip-2pt.11   &   -24$^{\rm h}$19$^{\rm m}$02$^{\rm s}$\hskip-2pt.5   &   18.9   \\
OGLE-2016-BLG-0117                      &   18$^{\rm h}$13$^{\rm m}$23$^{\rm s}$\hskip-2pt.75   &   -24$^{\rm h}$02$^{\rm m}$23$^{\rm s}$\hskip-2pt.8   &   20.0   \\
OGLE-2016-BLG-0244                      &   18$^{\rm h}$14$^{\rm m}$03$^{\rm s}$\hskip-2pt.42   &   -25$^{\rm h}$57$^{\rm m}$00$^{\rm s}$\hskip-2pt.5   &   19.0   \\
OGLE-2016-BLG-0118/OGLE-2016-BLG-0136   &   18$^{\rm h}$14$^{\rm m}$40$^{\rm s}$\hskip-2pt.44   &   -23$^{\rm h}$05$^{\rm m}$00$^{\rm s}$\hskip-2pt.7   &   15.0   \\
OGLE-2016-BLG-0141                      &   18$^{\rm h}$14$^{\rm m}$17$^{\rm s}$\hskip-2pt.47   &   -26$^{\rm h}$07$^{\rm m}$34$^{\rm s}$\hskip-2pt.4   &   19.1   \\
OGLE-2016-BLG-0089                      &   18$^{\rm h}$14$^{\rm m}$24$^{\rm s}$\hskip-2pt.48   &   -24$^{\rm h}$57$^{\rm m}$08$^{\rm s}$\hskip-2pt.7   &   21.3   \\
OGLE-2016-BLG-0086                      &   18$^{\rm h}$16$^{\rm m}$38$^{\rm s}$\hskip-2pt.93   &   -26$^{\rm h}$05$^{\rm m}$26$^{\rm s}$\hskip-2pt.4   &   13.1   \\
OGLE-2016-BLG-0193                      &   18$^{\rm h}$18$^{\rm m}$17$^{\rm s}$\hskip-2pt.63   &   -25$^{\rm h}$44$^{\rm m}$20$^{\rm s}$\hskip-2pt.9   &   18.4   \\
OGLE-2016-BLG-0111                      &   18$^{\rm h}$19$^{\rm m}$12$^{\rm s}$\hskip-2pt.07   &   -26$^{\rm h}$49$^{\rm m}$52$^{\rm s}$\hskip-2pt.2   &   17.4   \\
OGLE-2016-BLG-0114                      &   18$^{\rm h}$20$^{\rm m}$26$^{\rm s}$\hskip-2pt.72   &   -27$^{\rm h}$23$^{\rm m}$24$^{\rm s}$\hskip-2pt.6   &   18.6
\enddata
\tablenotetext{a}{Coordinates and magnitudes are taken from the OGLE Early Warning System (see \S \ref{sec:ground_surveys}).}
\tablenotetext{b}{Events are ordered by increasing R.A.}
\tablenotetext{c}{Events are named according to the following convention: Survey-Year-Field-Index, where the field designation ``BLG'' refers to the Galactic bulge.}
\tablenotetext{d}{For events discovered by multiple surveys, the event is named according to which survey alerted it first.}
\tablenotetext{e}{This includes the flux of the source as well as any blended stars.}
\label{tab:c9a_ongoing_events}
\end{deluxetable*}

Finally, we note that covering the {\ktcn} superstamp with ground-based surveys requires observations of a larger area than the area of the {\ktcn} superstamp, given the gaps between the {\kt} channels (e.g., near ${\rm RA} = 270.5^{\circ}$).
Any microlensing events and other time-variable sources that are not seen by {\kt} but that are within the areas covered by the ground-based surveys will have high-cadence multi-wavelength coverage, allowing in-depth study.
We also note that the superstamp area has been observed by microlensing surveys for many years and that many variable stars have been catalogued.
Specifically, there are 2140 RR Lyr stars \citep{soszynski2011}. 
Such information about variable stars will be used to improve photometry of microlensing events and will allow independent studies that are not the primary science driver for the {\ktcn} microlensing experiment.

%------------------------------------------------------------------------------------------------------------------------
\subsection{Photometric Methodology} \label{sec:k2_photometry}
%------------------------------------------------------------------------------------------------------------------------

\begin{figure}
   \centerline{
      \includegraphics[width=9cm]{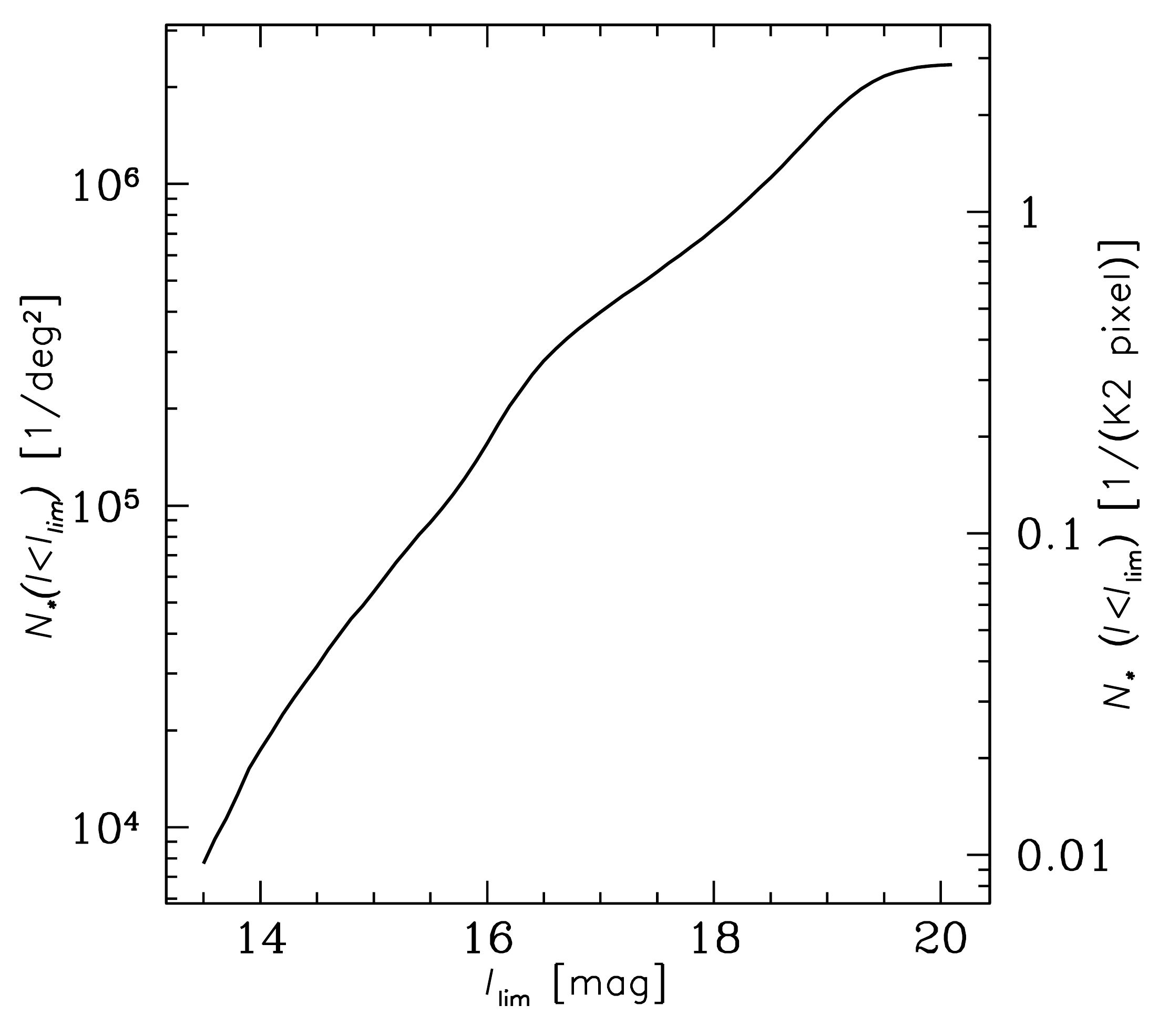}
   }
   \caption{
      \footnotesize{
         The cumulative distribution function of the number of stars brighter than a given limiting magnitude per square degree (left axis) and per {\kt} pixel (right).
         The data were taken from the OGLE-III maps of \citet{szymanski2011} for an $8\arcmin \times 17\arcmin$ subfield centered on (RA, Dec) = $(269.225^{\circ}, -28.3681^{\circ})$.
         The plot extends to $I\sim20$, at which there are $\sim$3 stars, on average, per {\kepler} pixel.
         This underscores one of the many challenges of performing crowded-field photometry toward the Galactic bulge with {\kt} data.
         \vspace{1.0mm}
      }
   }
   \label{fig:stardensity_cdf}
\end{figure}

Potentially the most important task of the Microlensing Science Team (MST) is to develop the tools necessary to extract photometry from the {\ktcn} data.
Accurate photometry of faint stars in very crowded fields must be measured from {\kt} images that have large pixels (4$\arcsec$), a PSF that is poorly sampled and at some level variable, and a non-uniform intra-pixel response.
Figure \ref{fig:stardensity_cdf} underscores this with a cumulative distribution function of the stellar density in one of the regions central to the {\ktcn} superstamp, where there are $\sim$3 stars per Kepler pixel with $I \lesssim 20$.
These problems are further complicated by the drift of stars across the focal plane that is caused by the torque of the Solar wind and is of-order 1 pixel/6 hours.

The majority of the MST is working on some aspect of the photometry problem, with support from both their proposed co-investigators and also volunteers.
Members of the MST will develop several photometric pipelines to process the {\ktcn} data.
The first, relying on difference imaging software designed for ground-based images, will provide quick-look photometry with no reliance on additional data.
A second technique will construct difference images for individual pixels and then fit the light curve parameters.
Thirdly, forward model difference imaging will use ground-based images and/or photometric catalogs to enable accurate modeling of the {\ktcn} images.
We will briefly describe the plans for all pipelines here, which are being developed and tested on data taken during \cz\ of the dense open cluster NGC 2158.

Difference image analysis (DIA; \citealt{tomaney1996,alard1998}) uses convolution to match the PSF of a high-quality reference image to that of a generally poorer-quality target image in order to enable the reference image to be subtracted from the target image, leaving only variable objects with non-zero flux in the residuals that comprise the difference image.
It is typically used on well-sampled ground-based images for which the variations in the PSF are caused by time-dependent changes in the seeing.
The {\kt} data differ significantly from the usual application of DIA in that the PSF does not differ significantly over the entire data set, and that the PSF is severely undersampled, suggesting that without significant modification, standard DIA software packages might not work well on {\kt} data.
Early experiments with \cz\ data of NGC 2158 seem to confirm this.
However, Penny \& Stanek (2016, in prep) showed that by first convolving the {\kt} images with a Gaussian in order to produce a well-sampled PSF, an algorithm that pairs the ISIS package \citep{alard1998,alard2000} with rudimentary detrending against pointing shifts could produce photometry of quality sufficient to detect and measure the variability of almost all known variable stars in the NGC 2158 cluster, including, most importantly, those with magnitudes and amplitudes similar to the microlensing events that will be observed in {\ktcn}.
The {\ktcn} pipeline based on this method will produce a light curve for every pixel in the {\ktcn} superstamp, as each pixel will contain several stars (see Figure \ref{fig:stardensity_cdf}).
Quick-look light curves of known microlensing events will be delivered to the Exoplanet Follow-up Observing Program (ExoFOP) site (see \S \ref{sec:community_exofop}) as soon as they are processed.
A full catalog of light curves for all pixels will be hosted on the NASA Exoplanet Archive at a date to be determined once the pipeline's performance has been evaluated.

A second difference imaging technique being investigated involves extending the detrending procedure on resolved stars to individual pixels, or combinations of pixels.
Applying the procedure to pixel-by-pixel light curves will generate a series of residual images.
These can be used in a similar fashion to classical difference images in that they can be used to identify the locations of variable stars, including microlensing events.
One approach to extracting photometry for variable stars is to combine the modeling of the intrinsic shape of the light curve profile with the detrending procedure.
By iteration, it is thus possible to find an optimal set of light curve parameters.
This method has been applied to eclipsing binaries within NGC 2158 using a simple model for the shape of the eclipse profiles.
Preliminary results have yielded a photometric precision of a few millimag for 15th-magnitude eclipse eclipsing binaries, and this same approach can be taken for microlensing events.

A final methodology involves forward modeling the {\kt} images using either star catalogs or higher-resolution ground-based images as the input.
Forward modeling refers to the process of producing a generative model of the {\kt} data by modeling the process by which stars cause charge to be collected in {\kepler}'s pixels.
For our purposes, this will involve the production of a model image for each 30-minute image from a model of the {\kepler} pixel response function (PRF) and a set of input data.
This model image will then be subtracted from the actual data, enabling photometry of the much less crowded variable sources that remain in the subtracted image.
For each image the {\kepler} PRF (as a function of wavelength and position) will be fit for (using well measured {\kepler} PRFs as strong priors) together with the pointing, roll, and distortion of the focal plane in order to produce each model image.
If, rather than a catalog, ground-based images are used for the input, a convolution kernel that matches the ground-based PSF to {\kepler}'s will be fit for instead of the PRF itself.
Because the pipeline for this method will rely on other data and the technique is new and will likely require more computing power, we expect the outputs of the pipeline to be delayed relative to the DIA-based pipeline.
However, we expect the photometry from this pipeline to improve significantly upon that from the DIA pipeline, enabling many additional microlensing events to be detected and measured from {\kt}.

%%%%%%%%%%%%%%%%%%%%%%%%%%%%%%%%%%%%%%%%%%%%%%%%%%
%%%
\section{Concurrent Ground-based Resources} \label{sec:ground}
%%%
%%%%%%%%%%%%%%%%%%%%%%%%%%%%%%%%%%%%%%%%%%%%%%%%%%

\begin{figure*}
   \centerline{
      \includegraphics[width=18cm]{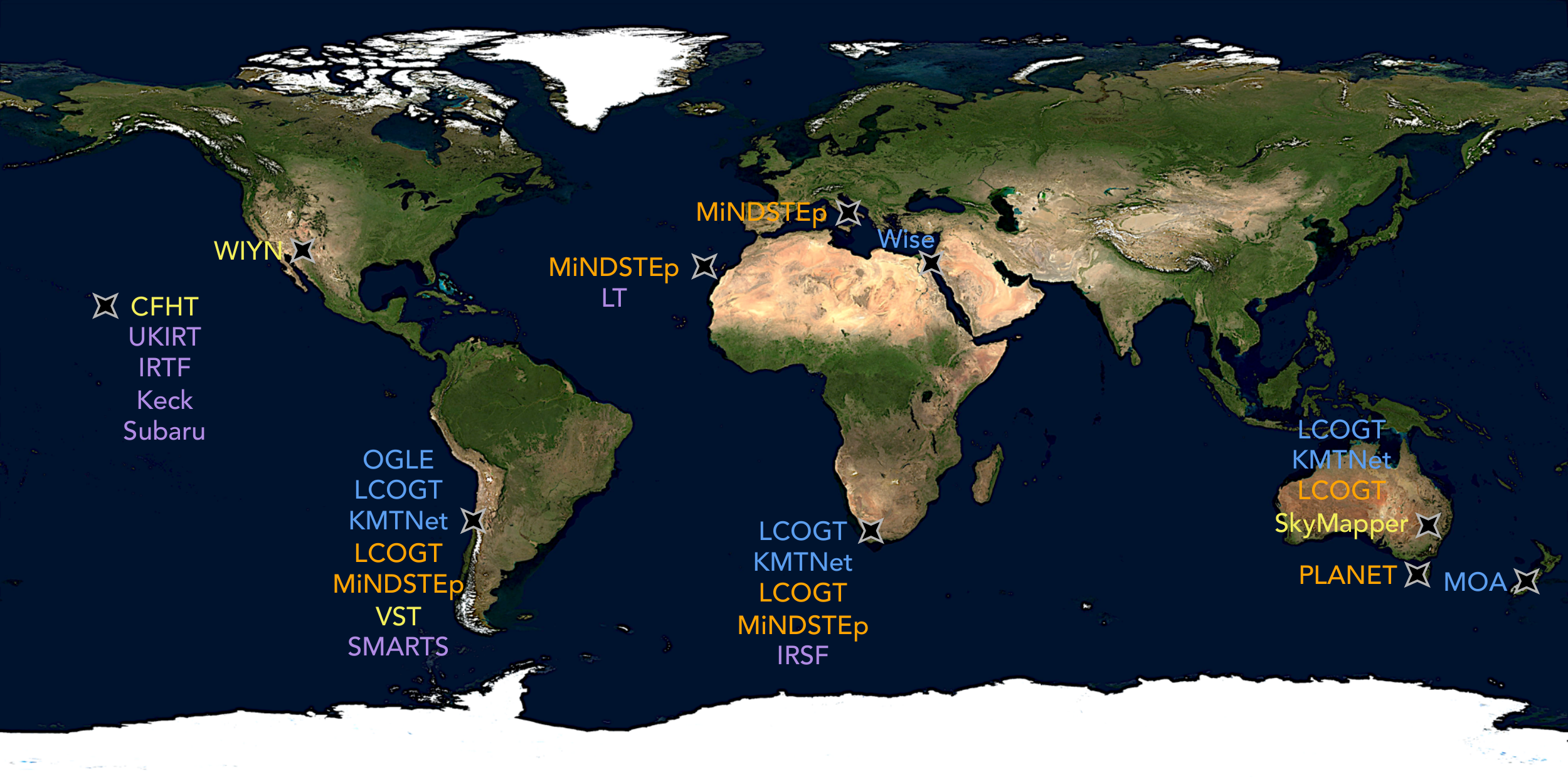}
   }
   \caption{
      \footnotesize{
         A map of all ground-based telescopes that the MST and others have procured to observe during {\ktcn}.
         Each is color-coded according to its primary scientific goal: automated survey (blue), high-cadence follow-up (orange), multiband photometric monitoring (yellow), and NIR source flux measurement (purple).
         Such a concerted effort will help to optimize the scientific return of {\ktcn}.
         \vspace{1.0mm}
      }
   }
   \label{fig:obs_map}
\end{figure*}

The MST and many members of the larger exoplanetary microlensing community have worked to secure a substantial network of ground-based resources that will observe in concert with {\ktcn}.
We broadly classify them according to four primary scientific motivations --- automated survey, high-cadence follow-up, multiband photometric monitoring, and NIR source flux measurement --- and discuss each in greater detail below.
Figure \ref{fig:obs_map} provides a map of the contributing observatories and Table \ref{tab:ground_resources} lists the parameters of each facility.
It is important to note that the specifications of available resources and their exact observing plans are subject to modification prior to the start of {\ktcn}; the final version of this paper will contain a more accurate accounting.
We then conclude with a discussion of the value of and efforts for real-time modeling of microlensing events during {\ktcn}.

%------------------------------------------------------------------------------------------------------------------------
\subsection{Automated Survey} \label{sec:ground_surveys}
%------------------------------------------------------------------------------------------------------------------------

The OGLE survey has been monitoring the Galactic bulge for microlensing phenomena for the last 24 years.
Since the discovery of the first microlensing event toward the Galactic bulge \citep{udalski1993}, OGLE has detected over 17,000 microlensing phenomena.
The vast majority of them were alerted to the community via the OGLE Early Warning System\footnote{\url{http://ogle.astrouw.edu.pl/ogle4/ews/ews.html}} \citep{udalski1994}.
In its current fourth phase, the OGLE-IV survey discovers over 2,000 real-time microlensing events annually, which constitutes about 90\% of lensing events toward the bulge.
The OGLE-IV facilities are located at the Las Campanas Observatory (LCO) in Chile.
The 1.3m Warsaw telescope and 256 Megapixel, 32-CCD detector mosaic camera, which covers 1.4 square degrees, have been used by OGLE-IV since 2010.
In 2016 OGLE-IV will continue its extensive monitoring of the Galactic bulge fields, adjusting somewhat the observing strategy to maximize coverage of the {\ktcn} superstamp.
Also, considerable effort will be undertaken to detect and alert a significant number of microlensing events outside the main superstamp region, which will provide targets for follow-up resources as well as candidates for ongoing event additions to the {\ktcn} target list (see \S \ref{sec:k2_pixels} and Tables \ref{tab:k2c9_parms} and \ref{tab:c9a_ongoing_events}).

MOA has similarly spent over a decade monitoring of the Galactic bulge to detect exoplanets via microlensing.
The second generation of MOA, MOA-II, is a 1.8m telescope with a 2.2 deg$^{2}$ FoV located at Mt.\ John University Observatory (MJUO) in New Zealand.
It will continue its concerted effort to reduce data daily and publish and circulate alerts of new microlensing events through their Transient Alert System\footnote{\url{https://it019909.massey.ac.nz/moa/}}.
Both OGLE-IV and MOA-II will observe the entire {\ktcn} superstamp with a cadence that is $\lesssim$1 hour.
Each will conduct their survey in a primary filter, $I$ for OGLE-IV and MOA-red for MOA-II, with occasional observations in $V$ for both surveys for source color measurements.
Table \ref{tab:ground_resources} provides a detailed list of the parameters of each facility.

The Las Cumbres Observatory Global Telescope (LCOGT) network, which consists of multiple telescopes at several northern and southern hemisphere sites, will also perform survey-mode operations during {\ktcn}.
At each of CTIO, SAAO, and SSO they expect to have equipped one 1.0m telescope with a $26\arcmin \times 26\arcmin$ Sinistro imager to provide survey capabilities at a wider range of sites and longitudes.
Wise Observatory in Israel will operate the Jay Baum 0.71m telescope, with a 1.0 square-degree FoV, to cover the {\ktcn} superstamp with 6 fields at a cadence of $\sim$30 minutes, and will use an Astrodon exoplanet filter that blocks light with $\lambda < 5000$\AA.
Finally, KMTNet will tile a substantial fraction of the {\ktcn} superstamp in $I$-band with a cadence of $\sim$10 minutes.
Together, OGLE-IV, MOA-II, LCOGT, and KMTNet will provide the dense coverage necessary to detect microlensing events and, in the case of OGLE-IV and MOA-II, generate and circulate alerts for new events on a $\sim$daily timescale.
Such information is not only crucial for constructing a database of known microlensing events within the {\ktcn} superstamp but also for providing real-time updates to targeted follow-up groups (see \S \ref{sec:ground_followup}) and quick-look photometry for real-time modeling analysis (see \S \ref{sec:realtime_modeling}), which itself is also useful for follow-up observations across all wavelengths.

%------------------------------------------------------------------------------------------------------------------------
\subsection{High-cadence Follow-up} \label{sec:ground_followup}
%------------------------------------------------------------------------------------------------------------------------

Although the current generation of microlensing surveys are able to observe many square degrees at an $\lesssim$hourly cadence, there are many advantages of collecting yet higher-cadence follow-up photometry of individual events.
The first is for event characterization.
While survey groups are indeed able to detect events as well as the perturbations induced by the presence of a planet, observing at a rate of several times more frequently can provide the most robust interpretation of the lens system, particularly in the case of high-magnification events \citep{griest1998,yee2012,han2013,yee2014,gould2014b}.
Furthermore, a higher cadence is optimal for securely detecting higher-order effects in light curves, including orbital and terrestrial parallax (see \S \ref{sec:ulens_parallax}) as well as orbital motion in the lensing system \citep{dominik1998,albrow2000,penny2011,shin2011,skowron2011,jung2013}, which causes the location and morphology of the caustics to change as a function of time.
A final benefit is that smaller-aperture smaller-FoV facilities can obtain observations if time is anticipated to be lost due to technical problems or weather for a survey telescope at a similar longitude.

\clearpage
\begin{turnpage}
\centering
\begin{deluxetable}{lccccccc}
\tablecaption{Ground-based Observing Resources Concurrent with {\ktcn}}
\tablewidth{0pt}
\tablehead{
\colhead{Strategy}   &
\colhead{Group/Telescope}   &
\colhead{Site}   &
\colhead{Aperture}   &
\colhead{FoV}   &
\colhead{Filter(s)}   &
\colhead{Cadence}   &
\colhead{Availability$^{a}$}   \\
\colhead{}   &
\colhead{}   &
\colhead{}   &
\colhead{[m]}   &
\colhead{}   &
\colhead{}   &
\colhead{[obs/unit time]}   &
\colhead{({\ktcn}: 7/April--1/July)}
}
\startdata
\textbf{Automated Survey}   \\
   &   OGLE   &   LCO   &   1.3   &   1.4 deg$^{2}$   &   $I$   &   20 minutes   &   7/April--1/July$^{b}$   \\
   &      &      &      &      &   $V$   &   1--2/night   &   \\
   &   MOA   &   MJUO   &   1.8   &   2.2 deg$^{2}$   &   MOA-red$^{c}$   &   15--90 minutes   &   7/April--1/July$^{b}$   \\
   &      &      &      &      &   $V$   &   nightly   &   \\
   &   LCOGT   &   CTIO   &   1.0   &   $26\arcmin \times 26\arcmin$   &   $i$   &   nightly   &   7/April--1/July   \\
   &      &   SAAO   &   1.0   &   $26\arcmin \times 26\arcmin$   &   $i$   &   nightly   &   7/April--1/July   \\
   &      &   SSO   &   1.0   &   $26\arcmin \times 26\arcmin$   &   $i$   &   nightly   &   7/April--1/July   \\
   &   Wise   &   Wise Observatory   &   0.71   &   1.0 deg$^{2}$   &   Astrodon$^{d}$   &   $\sim$30 minutes   &   7/April--1/July   \\
   &   KMTNet   &   CTIO   &   1.6   &   4.0 deg$^{2}$   &   $I$   &   $\sim$10 minutes   &   7/April--1/July$^{b}$   \\
   &      &   SAAO   &   1.6   &   4.0 deg$^{2}$   &   $I$   &   $\sim$10 minutes   &   7/April--1/July$^{b}$   \\
   &      &   SSO   &   1.6   &   4.0 deg$^{2}$   &   $I$   &   $\sim$10 minutes   &   7/April--1/July$^{b}$   \\
\textbf{High-cadence Follow-up}   \\
   &   LCOGT   &   CTIO   &   1.0 (2)$^{e}$   &   $15\arcmin \times 15\arcmin$   &   $gi$   &   ---   &   7/April--1/July   \\
   &      &   SAAO   &   1.0 (2)$^{e}$   &   $15\arcmin \times 15\arcmin$   &   $gi$   &   ---   &   7/April--1/July   \\
   &      &   SSO   &   1.0   &   $15\arcmin \times 15\arcmin$   &   $gi$   &   ---   &   7/April--1/July   \\
   &      &   SSO   &   2.0   &   $10\arcmin \times 10\arcmin$   &   $i$   &   ---   &   7/April--1/July   \\
   &   LT   &   La Palma   &   2.0   &   $10\arcmin \times 10\arcmin$   &   $gi$   &   ---   &   7/April--1/July   \\
   &   SMARTS   &   CTIO   &   1.3   &   $6.3\arcmin \times 6.3\arcmin$   &   $I$   &   ---   &   7/April--1/July   \\
   &      &      &      &      &   $V$   &   ---   &      \\
   &   IRTF   &   Mauna Kea   &   3.0   &   $60\arcsec \times 60\arcsec$   &   $I$   &   ---   &   20--25/May, 28/May--3/June,   \\
   &      &      &      &      &      &      &   6--10, 14--20, 24--26, 28/June   \\
   &   MiNDSTEp   &   La Silla   &   1.54   &   $45\arcsec \times 45\arcsec$   &   $VI$   &   ---   &   15/April--1/July   \\
   &      &   Fisciano   &   0.6   &   $22\arcmin \times 22\arcmin$   &   $I$   &   ---   &   7/April--1/July   \\
   &      &   SAAO   &   1.2   &   $13\arcmin \times 13\arcmin$   &   $I$   &   ---   &   7/April--1/July   \\
   &      &   Teide   &   1.0   &   $40\arcsec \times 40\arcsec$   &   $I$   &   ---   &   7/April--1/July   \\
   &   PLANET   &   UTGO   &   1.3   &   $20\arcmin \times 20\arcmin$   &   $VI$   &   ---   &   7/April--1/July   \\
\textbf{Multiband Photometric}   \\
\textbf{Monitoring}   \\
   &   CFHT   &   Mauna Kea   &   3.6   &   1.0 deg$^{2}$   &   $gri$   &   3 hours   &   TBD   \\
   &   SkyMapper   &   SSO   &   1.3   &   5.6 deg$^{2}$   &   $griz$   &   2--3 hours   &   7/April--1/July   \\
   &   WIYN   &   Kitt Peak   &   3.5   &   0.53 deg$^{2}$   &   $i$   &   30 minutes   &   9--14/May, 27/May--4/June,   \\
   &      &      &      &      &      &      &   5--9/July   \\
   &      &      &      &      &   $r$   &   nightly   &      \\
   &   VST   &   Cerro Paranal   &   2.6   &   1.0 deg$^{2}$   &   $Vr$   &   $\sim$3/night   &   7/April--1/July   \\
\textbf{NIR Source Flux}   \\
\textbf{Measurement}   \\
   &   UKIRT   &   Mauna Kea   &   3.8   &   0.2 deg$^{2}$   &   $H$   &   2--3/night   &   7/April--1/July   \\
   &   SMARTS   &   CTIO   &   1.3   &   $2.4\arcmin \times 2.4\arcmin$   &   $H$   &   targeted   &   7/April--1/July   \\
   &   IRSF   &   SAAO   &   1.4   &   $7.7\arcmin \times 7.7\arcmin$   &   $JHK^{f}$   &   TBD   &   TBD   \\
   &   LT   &   La Palma   &   2.0   &   $6\arcmin \times 6\arcmin$   &   $H$   &   targeted   &   7/April--1/July   \\
   &   IRTF   &   Mauna Kea   &   3.0   &   $60\arcsec \times 60\arcsec$   &   $H$   &   targeted   &   20--25/May, 28/May--3/June,   \\
   &      &      &      &      &      &      &   6--10, 14--20, 24--26, 28/June   \\
   &   Keck   &   Mauna Kea   &   10.0   &   $40\arcsec \times 40\arcsec$   &   $H$   &   4 ToO triggers   &   7/April--1/July$^{g}$   \\
   &   Subaru   &   Mauna Kea   & 8.2   &   $21\arcsec \times 21\arcsec$   &   TBD   &   targeted (2 hours)   &   24/June
\enddata
\tablenotetext{a}{Only dates during {\ktcn} are considered here, though many contributing facilities may observe within the {\ktcn} superstamp prior to and following {\cn}.}
\tablenotetext{b}{Observations of the {\ktcn} superstamp will occur during the entire bulge observing season, guaranteeing coverage of events that peak before and after {\cn}.}
\tablenotetext{c}{The MOA-red filter is well approximated by $R$+$I$.}
\tablenotetext{d}{This filter blocks light with $\lambda < 5000$\AA.}
\tablenotetext{e}{LCOGT will operate multiple telescopes with the specified aperture at the specified site.}
\tablenotetext{f}{IRSF will take data in all three filters ($JHK$) simultaneously.}
\tablenotetext{g}{The Keck ToOs are only guaranteed possible during Caltech and U of C system allocations on Keck 2.}
\label{tab:ground_resources}
\end{deluxetable}
\end{turnpage}
\clearpage

To this effect LCOGT will operate two 1.0m telescopes each at CTIO and SAAO and one 1.0m at SSO, all with a  $15\arcmin \times 15\arcmin$ FoV, as well as a 2.0m telescope at SSO with a  $10\arcmin \times 10\arcmin$ FoV.
These will produce concentrated $i$-band observations of selected events during {\ktcn}.
The PLANET collaboration will operate the 1.3m Harlingten telescope at the University of Tasmania, Greenhill Observatory (UTGO).
During the campaign it will be primarily dedicated to follow-up of {\ktcn} microlensing targets in $V$ and $I$ with a $20\arcmin \times 20\arcmin$ FoV camera.
The observations will be coordinated with other facilities operated by LCOGT and the MST.
PLANET will provide real-time photometry of the observed microlensing events and alerts for potential anomalies.
MiNDSTEp will contribute continuous high-cadence extended $V$ and $I$-band observations from the Danish 1.54m telescope at ESO's La Silla observatory in Chile, equipped with a two-color EMCCD lucky imaging camera with a $45\arcsec \times 45\arcsec$ FoV, operated at 10 Hz time resolution \citep{skottfelt2015}, and continuous high-cadence $I$-band observations from the Salerno University 0.6m telescope located in Fisciano, Italy, equipped with a CCD camera with a $22\arcmin \times 22\arcmin$ FoV.
MiNDSTEp also expects to provide $I$-band observations from the MONET-South 1.2m telescope at SAAO in South Africa, equipped with a back-illuminated CCD camera and a FoV of $12.6\arcmin \times 12.6\arcmin$, and from the 1m SONG Hertzsprung telescope at Tenerife, equipped with an EMCCD lucky imaging camera with a $40\arcsec \times 40\arcsec$ FoV.
Additionally, the SMARTS 1.3m telescope at CTIO, the 2.0m Liverpool Telescope (LT) at La Palma, and the 3.0m Infrared Telescope Facility (IRTF) at Mauna Kea, the primary purpose of all of which is for NIR source flux measurement for short-timescale events (see \S \ref{sec:ground_nir}), will take observations in optical bands simultaneous with their NIR flux measurements, as each is equipped with a dual-channel optical+NIR imager.
The parameters of each of these resources is included in Table \ref{tab:ground_resources}.

%------------------------------------------------------------------------------------------------------------------------
\subsection{Multiband Photometric Monitoring} \label{sec:ground_color}
%------------------------------------------------------------------------------------------------------------------------

Multiband time-series photometry plays two critical roles for {\ktcn}, and in both cases the goal of the observations is to measure the color and magnitude of the source star.
Firstly, knowledge of the source color will be invaluable in measuring all parameters of the microlensing events seen from {\kepler}.
This is because the events seen by {\kepler} will be highly blended, and will be in the so-called pixel lensing regime \citep{crotts1992,baillon1993}, wherein the impact parameter {\uzero}, timescale {\tE}, and the source flux relative to blended light are strongly degenerate (see \citealt{riffeser2006} for an overview).
The impact parameter must be measured in order to determine the parallax, so if the degeneracy is not broken it is only possible to measure one-dimensional parallaxes \citep{gould2014a}.
However, if the source magnitude in the {\kepler} bandpass can be inferred from ground-based monitoring of the event in one or more filters, the impact parameter for {\kepler} can be better constrained.
{\kepler}'s bandpass is broad, covering ${\sim}430$--$880$~nm, and so covers $BVRI$ or $griz$ bandpasses, though with only partial overlap of $B$, $g$, and $z$.
Reconstruction of the source's {\kepler} magnitude $K_{p}$ can best be done with knowledge of the source magnitude in several filters that cover the {\kepler} bandpass (especially in regions where there is significant differential extinction), but it can also be achieved to lesser accuracy with just a single color.

The second application of a source color is the measurement of the source's angular diameter through a color-angular diameter relation.
This becomes important if the microlensing light curve displays finite-source effects, as this allows the conversion of these effects into a measurement of the angular Einstein ring radius {\thetaE} \citep{yoo2004}, which together with microlens parallax {\piE} can be used to fully solve the event and measure the lens mass {\ml} and the lens-source relative parallax {\pirel} (see \S \ref{sec:ulens_parallax}).
The angular diameter measurements are best made using the widest practical wavelength baseline; $V-I$ has sufficient baseline and is regularly used in practice, and $r-z$ has a similar baseline but may be more useful in regions of high extinction.
If NIR measurements are possible, then visual minus NIR colors can be used and may prove to be more accurate.

In all cases, measurements of the source color must be made using time-series photometry in order to separate the varying, magnified source flux from any blended light whose magnitude is constant in time.
For short-timescale FFP events, there may not be enough time to alert follow-up observations to obtain multicolor observations, so it is necessary to survey the entire {\cn} superstamp with a cadence of at least a few hours in each filter in order to ensure the color measurements are possible.

There are several facilities that will contribute the aperture and FoV necessary to obtain multiband photometric monitoring across the {\ktcn} superstamp.
Table \ref{tab:ground_resources} details the parameters of each of them.
The OGLE and MOA surveys will obtain occasional $V$-band data (see Table \ref{tab:ground_resources}), which will provide source color measurements for some events.
But, as explained above, it is important to measure the source color for all events, regardless of timescale or magnitude, and to do so in multiple filter combinations and across long wavelength baselines.
The Canada France Hawai'i Telescope (CFHT) on Mauna Kea, with a 3.6m aperture and 1.0 deg$^{2}$ FoV, will take $gri$ data twice per night.
SkyMapper, a 1.3m telescope with a 5.6 deg$^{2}$ FoV located at SSO, will cycle through $griz$ every 2--3 hours.
The one-degree imager on the WIYN 3.5m at Kitt Peak will take $r$- and, less frequently, $i$-band images.
Lastly, the 2.6m ESO-operated VLT Survey Telescope (VST), which has a 1.0 square-degree FoV and is located in Cerro Paranal, will contribute $V$ and $r$ observations $\sim$3 times per night.
However, even with the involvement from all of these observatories, telescope and instrument scheduling means that the color coverage is not complete over the entire campaign.
There is thus a significant role to be played by follow-up observations.
Additionally, there will be no survey-style multiband coverage for events outside the superstamp, so color follow-up observations are essential for interpretation of events monitored by {\kt} outside the superstamp.

%------------------------------------------------------------------------------------------------------------------------
\subsection{NIR Source Flux Measurement} \label{sec:ground_nir}
%------------------------------------------------------------------------------------------------------------------------

By tiling the {\ktcn} superstamp with NIR facilities it will be possible to determine the NIR source flux of most if not all microlensing events.
As discussed in detail in \S \ref{sec:ulens_flux}, such an effort will provide a second method by which to directly measure ({\mstar}, {\mpl}, {\dl}) that is independent from {\piE}.
{\ktcn}, then, will provide a large control sample with which we can refine and calibrate the flux characterization-derived results with satellite parallax values, which moreover is crucial pathfinding work in advance of {\wfirst} \citep{spergel2015}, as NIR flux characterization may be the dominant mechanism by which to derive the fundamental parameters of the planetary systems {\wfirst} will detect.\footnote{If concurrent ground-based observations are taken toward the {\wfirst} target fields, the satellite parallax method can also be used to derive planetary parameters \citep{yee2013,zhu2016}.}
Additionally, NIR source flux measurements are integral for ultimately deriving the strongest constraints possible on the nature of FFP candidates.

The United Kingdom Infrared Telescope (UKIRT), with a 3.8m aperture and a 0.20 deg$^{2}$ FoV, will conduct an automated survey of the {\ktcn} superstamp through the campaign with a cadence of 2--3 observations per night.
In principle this will be sufficient to measure the NIR source flux for all microlensing events save those with the very shortest timescales.
As these are characteristic of FFPs, one of the primary scientific drivers for {\ktcn}, the MST have worked to procure an array of NIR facilities able to trigger NIR follow-up for these events.
Specifically, the SMARTS 1.3m at CTIO, the IRTF at Mauna Kea, the LT at La Palma, and the 1.4m Infrared Survey Facility (IRSF) at SAAO will all target individual microlensing events to guarantee NIR source flux measurements.
Additionally, the 8.2m Subaru telescope, located at Mauna Kea, will contribute two hours of targeted follow-up in NIR bands on 24/June.

A final, experimental venture to this end is the use of NIRC2 on Keck to trigger target-of-opportunity (ToO) observations of hand-picked short-timescale events.
The MST was awarded four such ToO triggers during {\ktcn}, the goal being to obtain the first epoch of magnified NIR data for short-timescale FFP candidates described in \S \ref{sec:ulens_flux}.
It is true that all of the NIR resources that have hitherto been discussed are able to accomplish this task.
Nevertheless, as the second epoch must necessarily be taken with a high-resolution facility, taking the first epoch using the same instrument on the same telescope allows for the strongest possible lens flux constraints.
In Table \ref{tab:ground_resources} we provide a catalog of the parameters of all observatories.

%------------------------------------------------------------------------------------------------------------------------
\subsection{Real-time Modeling} \label{sec:realtime_modeling}
%------------------------------------------------------------------------------------------------------------------------

Along with the aggregation of telescopes listed above, a real-time modeling effort will be essential to the success of {\ktcn}.
While {\kt} itself and many of the ground-based facilities will operate in an automated fashion, all targeted data collection efforts, and in particular those contributing NIR imagers, will benefit from and rely on some form of real-time event analysis.
This capability helps efficaciously allocate resources to events with high observational and/or scientific priority via rapid interpretation of the temporal evolution of the events.
In specific, predictions by real-time modeling efforts help to predict caustic crossings in order to guarantee the dense observations necessary to constrain the microlensing observables (see \S \ref{sec:ulens_geometry}).
Furthermore, the rapid and robust determination of short-timescale microlensing events is crucial for any NIR facilities.
Having a well-developed modeling pipeline is of paramount importance for eliminating false positives and utilizing the Keck ToO triggers on the candidates most likely to yield secure FFP detections.
A prompt classification of anomalous events can also be useful for deploying additional ToO facilities for genuine planetary events, identifying stellar binary contaminants, and preventing the use of expensive facilities on less interesting events.

Within the microlensing community, there are several active groups providing real-time modeling of binary and planetary events.
These groups have developed their own codes using different algorithms that naturally provide independent checks for the proposed solutions.
The modeling of binary microlensing events is made particularly difficult by the existence of caustics (see \S \ref{sec:ulens_geometry}), which rapidly change their shapes for small variations in the parameters and may abruptly create peaks or dips in the light curves.
For this reason, many disconnected local minima for the chi-squared function can coexist in the parameter space.
With the purpose of making the exploration as exhaustive and fast as possible, two strategies have been proposed to set the initial conditions of downhill fitting: a grid search in the parameter space, or template-matching from a wide library of light curves \citep{mao1995,liebig2015}.
The latter is the strategy adopted by the fully automatic platform RTModel\footnote{\url{http://www.fisica.unisa.it/gravitationAstrophysics/RTModel.htm}}, which is able to provide predictions for the light curves as seen by {\kepler} using available ground-based observations.

%%%%%%%%%%%%%%%%%%%%%%%%%%%%%%%%%%%%%%%%%%%%%%%%%%
%%%
\section{Synergy Between {\ktcn} and {\spitzer} Microlensing} \label{sec:spitzer}
%%%
%%%%%%%%%%%%%%%%%%%%%%%%%%%%%%%%%%%%%%%%%%%%%%%%%%

\begin{figure*}
   \centerline{
      \includegraphics[width=18cm]{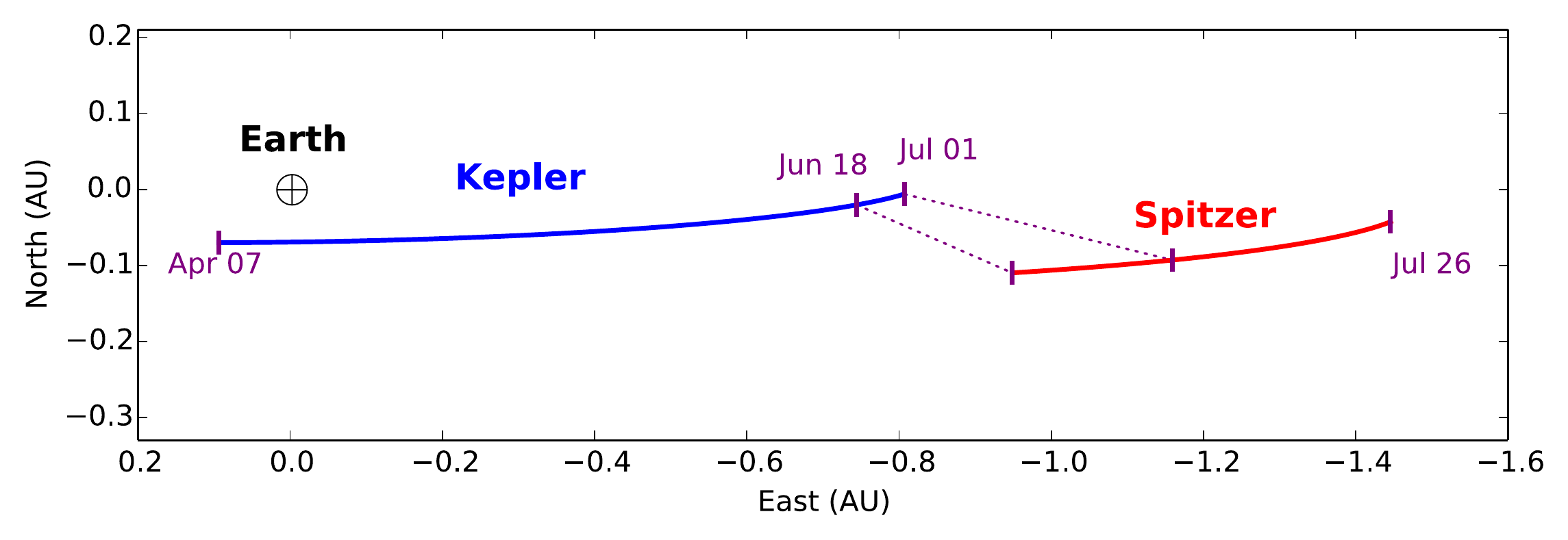}
   }
   \caption{
      \footnotesize{
         The projected positions of {\kepler} and {\spitzer} with respect to the Earth, as seen from the center of the {\ktcn} superstamp during their respective campaign periods.
         \vspace{1.0mm}
      }
   }
   \label{fig:k2_spitzer_positions}
\end{figure*}

From June 18 to July 26\ in 2016, {\spitzer} will also be able to observe the Galactic bulge, leading to a 13-day overlap with the {\ktcn} window.

\begin{figure}
   \centerline{
      \includegraphics[width=9cm]{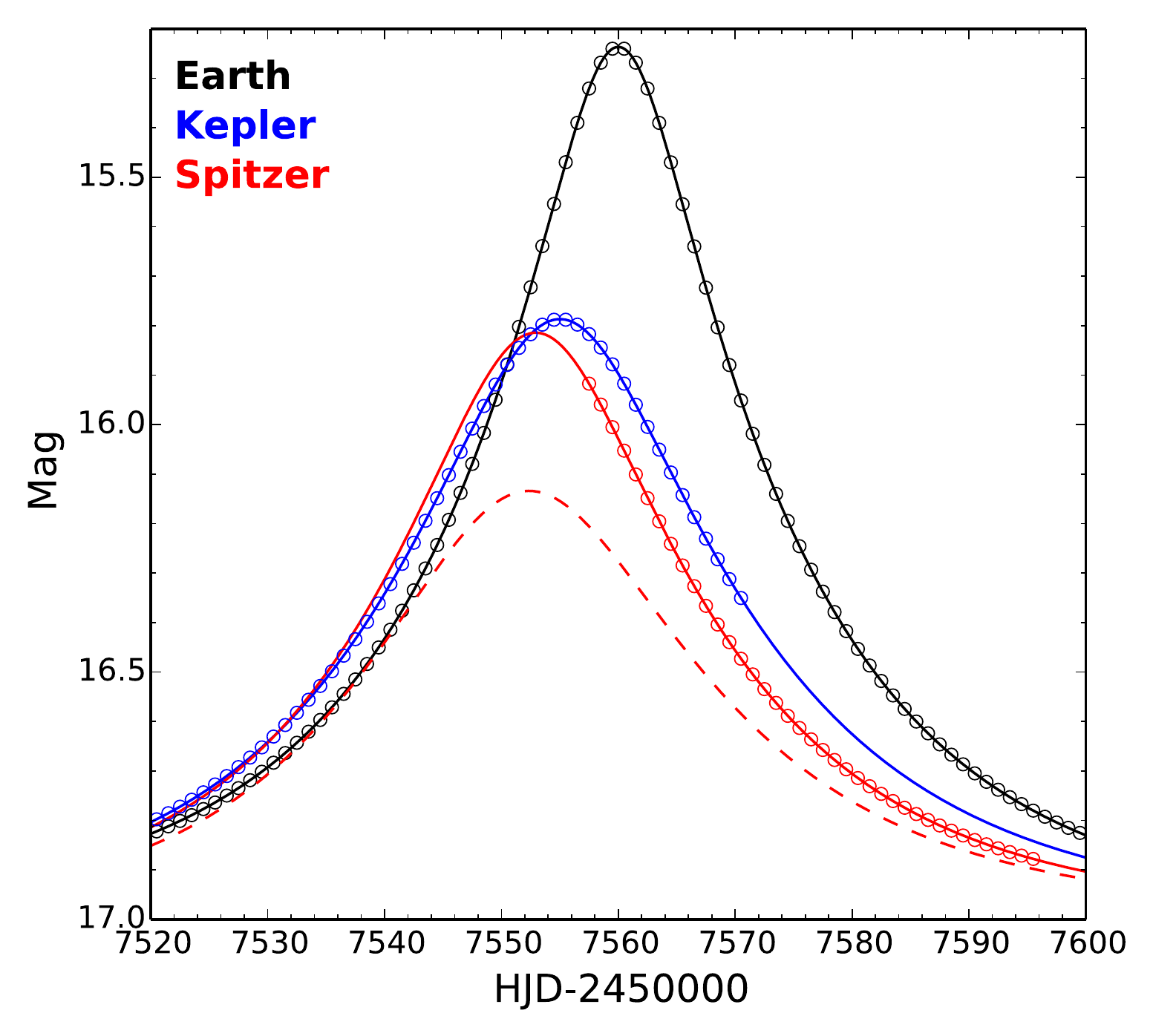}
   }
   \caption{
      \footnotesize{
         An example event in which the four-fold degeneracy can be broken by combining observations from the Earth, {\kepler}, and {\spitzer}.
         With observations from the Earth and {\kepler}, there are four allowed solutions of $\bdv{\pi}_{\rm E}$, which lead to two distinct predictions for the light curve as seen by {\spitzer}.
         The {\spitzer} observations can easily identify the correct solution.
         This event has typical parameters for disk lenses ($u_0=0.2$, $t_{\rm E}=30$ days, $\pi_{\rm E,N}=\pi_{\rm E,E}=0.2$).
         \vspace{1.0mm}
      }
   }
   \label{fig:k2_spitzer_break_degeneracy}
\end{figure}

Gould, Yee, and Carey have an accepted {\spitzer} program to conduct a two-satellite microlensing experiment (PI: A.\ Gould, \citealt{gould2015a}).
The primary goal of this {\kt} plus {\spitzer} endeavor is to demonstrate the idea of using an additional satellite to break the four-fold degeneracy that is present in the case of observations from a single-satellite (in addition to those from the ground) \citep{refsdal1966,gould1994a}.
In a single-satellite experiment (e.g., {\ktcn} or {\spitzer}), the microlens parallax vector is given by:
\begin{equation} \label{eq:pievec}
   \bdv{\pi}_{\rm E} = \frac{\rm AU}{D_\perp} \left(\frac{\Delta t_0}{t_{\rm E}},\ \Delta u_0\right).
\end{equation}
Here $\Delta t_0 \equiv t_{0,\rm sat}-t_{0,\oplus}$ and $\Delta u_0 \equiv u_{0,\rm sat}-u_{0,\oplus}$ are the differences in the peak times and impact parameters as seen from the two sites, respectively.
While the light curves can yield $t_{0,\rm sat}$ and $t_{0,\oplus}$ unambiguously, they can only yield the absolute values of impact parameters, $|u_{0,\rm sat}|$ and $|u_{0,\oplus}|$.
Hence, Equation~(\ref{eq:pievec}) is four-fold ambiguous (see, e.g., Figure 1 of \citealt{gould1994a}):
\begin{equation} \label{eq:4fold}
   \bdv{\pi}_{\rm E} = \frac{\rm AU}{D_\perp} \left(\frac{t_{0,\rm sat}-t_{0,\oplus}}{t_{\rm E}},\ \pm |u_{0,\rm sat}| \pm |u_{0,\oplus}|\right).
\end{equation}
This degeneracy typically leads to two distinct solutions for the lens mass and distance.
The four-fold degeneracy can be broken in specific cases, such as planetary events, high-magnification events, or events with kinematic information \citep{yee2015b}, and can be approached statistically for a sample of events \citep{calchinovati2015}.
However, it can only be systematically broken by obtaining observations from a second, misaligned, satellite (\citealt{refsdal1966,gould1994a}; see also \citealt{gaudi1997}).
The addition of {\spitzer} ({\kepler}) to {\kepler} ({\spitzer}) fulfills such a requirement, as is shown in Figure~\ref{fig:k2_spitzer_positions}.
Figure \ref{fig:k2_spitzer_break_degeneracy} shows the light curve for an event with parameters typical of a lens in the Galactic disk as seen by the Earth, {\kepler}, and {\spitzer}.
Observations with {\spitzer} can easily identify the correct solution, leading to the unique determination of {\piE}.
When combined with a measurement of {\thetaE} or the lens flux, this uniquely solves for the lens mass and distance.

An ensemble of single-lens events for which the four-fold degeneracy has been broken can be used to test the Rich argument.
Rich's argument, which asserts that the parallactic shift in impact parameter, $\Delta{\uzero}$, should be the same order as the parallax-induced shift in {\tzero}, $\Delta {\tzero}/{\tE}$, is used to statistically break the four-fold degeneracy \citep{calchinovati2015}.
For events with $\theta_{\rm E}$ measurements, the resolution of this four-fold degeneracy can directly yield precise mass and distance measurements of the lens system without requiring any additional arguments or observations \citep{zhu2015b}.
The inclusion of a second satellite can also break the 1-D continuous parallax degeneracy that can be present in events with a binary lens system.
Since $\theta_{\rm E}$ is nearly always measured in such cases, this leads to more precise measurement of the mass of the binary lens.

With 50 hours of {\spitzer} time, $\sim$25 events that fall inside the {\ktcn} superstamp are expected to be observed.
This subset will include several binaries that remain active when {\spitzer} observations begin, and $\sim$20 relatively bright single-lens events selected from a sample of $\sim$50 events that will peak within a 30-day window that is centered on June 24, the midpoint of the 13-day overlap window.
These events will follow the standard {\spitzer} event selection procedure \citep{udalski2015b}: they are selected based on ground-based observations, uploaded to the {\spitzer} spacecraft on Mondays, and observed starting the following Thursday.

%%%%%%%%%%%%%%%%%%%%%%%%%%%%%%%%%%%%%%%%%%%%%%%%%%
%%%
\section{Community Involvement} \label{sec:community}
%%%
%%%%%%%%%%%%%%%%%%%%%%%%%%%%%%%%%%%%%%%%%%%%%%%%%%

{\ktcn} is designed to provide access to a myriad of compelling science goals in a way that is community-driven.
One of the most critical components of this white paper is the description and dissemination of the opportunities for involvement for personnel outside of the field of exoplanetary microlensing.
Here we describe the online access to relevant data products.
It is our intention that this will encourage involvement in exoplanetary microlensing generally and {\ktcn} specifically and will help to maximize the scientific yield of {\ktcn}.

%------------------------------------------------------------------------------------------------------------------------
\subsection{ExoFOP Interface} \label{sec:community_exofop}
%------------------------------------------------------------------------------------------------------------------------

During the {\kepler} primary mission, the NASA Exoplanet Science Institute (NExScI) developed a website for coordination and collation of ground-based follow-up observation activities by the {\kepler} Science Team.
During the extended {\kt} mission, this site was transitioned for support of the entire {\kepler} community and renamed the Community Follow-up Observing Program (CFOP).
CFOP enables users to share images, spectra, radial velocities, stellar parameters, planetary parameters, observational parameters, free-form observing notes, false-positive alerts, and any type of file the users wish to upload.
Currently, CFOP contains over 100,000 files and 25,000 parameters on 7,500 {\kepler} objects of interest --- all uploaded by registered users and available for use by the community.
In 2015, CFOP was used as the basis for an expanded site (ExoFOP) to support the {\kt} mission, and will be used in the future to support TESS, NN-EXPLORE RV targets, and eventually {\wfirst} exoplanet (coronagraphic and microlensing) targets.
For {\kt}, ExoFOP includes all targets and users can upload the same types of files and data as above, and can designate target status such as `planet candidate,' `false positive,' or `eclipsing binary.'
To date, users have uploaded over 40,000 files and identified over 200 planet candidates.
CFOP and ExoFOP are developed and operated by NExScI with funding from the {\kepler} project (for CFOP) and from the NASA Exoplanet Archive (for ExoFOP).

As {\ktcn} will not be driven by pre-identified targets, ExoFOP support for the microlensing campaign will be specifically tailored.
The general strategy was designed in discussions with the MST.
The three main components will be:
\begin{enumerate}
   \item a sortable table containing all microlensing events identified within the {\ktcn} superstamp,
   \item detailed information for each event (e.g., cursory single-lens fit parameters and magnitudes), and
   \item a graphical display of available telescope resources.
\end{enumerate}
The event list will be driven by events collected by LCOGT's RoboNet \citep{tsapras2009}, which accrues events and photometry from the dedicated ground-based microlensing projects OGLE, MOA, and LCOGT.
Basic information about all events, such as preliminary real-time single-lens fit parameters ({\tzero}, {\tE}, {\uzero}) and current apparent magnitude, will be available in a single, sortable table, similar to the {\kt} campaign tables currently on ExoFOP.
The detailed information for each event will include quick-look photometry, images, and detailed real-time modeling results.
Information collated by RoboNet will be automatically available and users will also be able to upload data, model parameters, files, and free-form observing notes.
The telescope resources display will have a large-scale calendar version covering the full duration of {\ktcn}, as well as the ability to generate a detailed visualization of the observability of the {\ktcn} superstamp for each ground-based site for a single day.
The goal of these graphics is to help coordinate the timing of ground-based observations.
In addition, there will be a search interface covering all data and user notes.

The ExoFOP website\footnote{\url{https://exofop.ipac.caltech.edu/}} is open to the entire community.
In ExoFOP, all data and uploaded files are visible to all users.
To upload content, users must have an account and be logged in.
The same user account works on both CFOP and ExoFOP and a user account can be requested by following the link on the ExoFOP home page.

%------------------------------------------------------------------------------------------------------------------------
\subsection{{\ktcn} Visibility Tool} \label{sec:community_k2fov}
%------------------------------------------------------------------------------------------------------------------------

The \texttt{K2fov} tool \citep{mullally2016} allows users to check whether a list of input target coordinates will fall within the {\kt} FoV during a user-specified campaign.
This functionality has been expanded for {\ktcn} and is available through an in-browser application\footnote{\url{http://k2c9.herokuapp.com/}}.
Given that many of the teams representing ground-based resources will eschew a proprietary period for their data and will act to host the photometry on ExoFOP, this allows users to determine if a desired target will have publicly available data across a wide range of wavelengths and cadences.

%%%%%%%%%%%%%%%%%%%%%%%%%%%%%%%%%%%%%%%%%%%%%%%%%%
%%%
\acknowledgments
%%%
%%%%%%%%%%%%%%%%%%%%%%%%%%%%%%%%%%%%%%%%%%%%%%%%%%

CBH, RP, MP, RAS, DPB, DWH, and BSG were supported through the NASA {\kt} Guest Observer Program.
This research has made use of the NASA Exoplanet Archive, which is operated by the California Institute of Technology, under contract with the National Aeronautics and Space Administration under the Exoplanet Exploration Program.
Work by CBH and YS was supported by an appointment to the NASA Postdoctoral Program at the Jet Propulsion Laboratory, administered by Universities Space Research Association through a contract with NASA.
The OGLE project has received funding from the National Science Centre, Poland, grant MAESTRO 2014/14/A/ST9/00121 to AU.
GD acknowledges Regione Campania for support from POR-FSE Campania 2014-2020.
TCH is funded through KASI grant \#2016-1-832-01.
SM was supported by the Strategic Priority Research Program ``The Emergence of Cosmological Structures'' of the Chinese Academy of Sciences Grant No.\ XDB09000000, and by the National Natural Science Foundation of China (NSFC) under grant number 11333003 and 11390372 (SM).
CBH thanks graphic designer Kathryn Chamberlain for her generous assistance with Figure 4.

%%%%%%%%%%%%%%%%%%%%%%%%%%%%%%%%%%%%%%%%%%%%%%%%%%
%%%
\bibliographystyle{apj}
%%%
%%%%%%%%%%%%%%%%%%%%%%%%%%%%%%%%%%%%%%%%%%%%%%%%%%

\end{document}